\documentclass[11pt,a4paper]{article}
\pdfoutput=1

\usepackage{jheppub}
\usepackage{bm}

\usepackage[english]{babel}
\usepackage{slashed}
\usepackage{multirow}
\usepackage[normalem]{ulem}

\newcommand {\E}[1]{\times 10^{#1}}	
\newcommand {\e}[1]{\mathrm{~#1}}       
\newcommand{\mc}[1]{\mathcal{#1}}

\newcommand{\mrm}[1]{\mathrm{#1}}
\newcommand{\re}[0]{\mrm{Re}}
\newcommand{\im}[0]{\mrm{Im}}

\newcommand{\nn}[0]{\nonumber}

\newcommand{\pmns}[0]{V_\mrm{PMNS}}

\definecolor{Red}{rgb}{1.,0.,0.}

\definecolor{Green}{rgb}{0.2,.7,0.2}

\definecolor{nicered}{rgb}{0.7,0.1,0.1}
\definecolor{nicegreen}{rgb}{0.1,0.5,0.1}

\usepackage{hyperref}
\hypersetup{colorlinks,citecolor=nicegreen,linkcolor= nicered}

\begin{document}

\arxivnumber{1306.6493}

\author[a]{Ilja Dor\v sner} 
\author[b,c]{Svjetlana Fajfer} 
\author[c,d]{Nejc Ko\v snik} 
\author[c]{Ivan Ni\v sand\v zi\'c}

\affiliation[a]{Department of Physics, University of Sarajevo, Zmaja
  od Bosne 33-35, 71000 Sarajevo, Bosnia and Herzegovina}
\affiliation[b]{Department of Physics,
  University of Ljubljana, Jadranska 19, 1000 Ljubljana, Slovenia}
\affiliation[c]{J. Stefan Institute, Jamova 39, P. O. Box 3000, 1001
  Ljubljana, Slovenia}
\affiliation[d]{Institute of Metals and Technology, Lepi pot 11,
  Ljubljana, Slovenia}

\emailAdd{ilja.dorsner@ijs.si}
\emailAdd{svjetlana.fajfer@ijs.si}
\emailAdd{nejc.kosnik@ijs.si}
\emailAdd{ivan.nisandzic@ijs.si}

\title{Minimally flavored colored scalar in $\bar B \to D^{(*)} \tau
  \bar \nu$ and the mass matrices constraints}

\date{\today}

\abstract{
  The presence of a colored scalar that is a weak doublet
  with fractional electric charges of $|Q|=2/3$ and $|Q|=5/3$ with
  mass below $1$\,TeV can provide an explanation of the observed
  branching ratios in $B \to D^{(*)} \tau \bar \nu$ decays. The
  required combination of scalar and tensor operators in the effective
  Hamiltonian for $b \to c \tau \bar \nu$ is generated through the
  $t$-channel exchange. We focus on a scenario with a minimal set of
  Yukawa couplings that can address a semitauonic puzzle and show that
  its resolution puts a nontrivial bound on the product of the scalar
  couplings to $\bar \tau b$ and $\bar c \nu$. We also derive
  additional constraints posed by $Z \to b\bar b$, muon magnetic
  moment, lepton flavor violating decays $\mu \to e \gamma$, $\tau \to \mu \gamma$, $\tau \to e \gamma$, and $\tau$
  electric dipole moment. The minimal set of Yukawa couplings is not
  only compatible with the mass generation in an $SU(5)$ unification
  framework, a natural environment for colored scalars, but specifies
  all matter mixing parameters except for one angle in the up-type
  quark sector. We accordingly spell out predictions for the proton
  decay signatures through gauge boson exchange and show that $p
  \rightarrow \pi^0 e^+$ is suppressed with respect to $p \rightarrow
  K^+ \bar{\nu}$ and even $p \rightarrow K^0 e^+$ in some parts of
  available parameter space. Impact of the colored scalar embedding in
  $45$-dimensional representation of $SU(5)$ on low-energy
  phenomenology is also presented. Finally, we make predictions for
  rare top and charm decays where presence of this scalar can be
  tested independently.
}


\maketitle

\section{Introduction}
Results from LHC and $B$-factories have narrowed down the
possibilities for New Physics (NP) close to the electroweak scale. After
precise measurements of many observables in $\Delta B=2$ and $\Delta
B=1$ transitions hopes for the presence of NP persist only in
the sector of leptonic and/or semileptonic decays. Namely, there have been
experimental indications of enhanced branching ratios in the $B\to D(D^*) \tau
\bar\nu_\tau$ decays with respect to the Standard Model (SM) predictions. In particular, BaBar collaboration
has presented the following ratios~\cite{Babar}
\begin{eqnarray}
\mathcal R^*_{\tau/\ell} = {\mathcal B(B\to D^* \tau \nu)}/{\mathcal B (B\to D^{*}\ell \nu)} = 0.332 \pm 0.030\,, \label{eq:Rstar}\\
\mathcal R_{\tau/\ell} = {\mathcal B(B\to D \tau \nu)}/{\mathcal B(B\to D^{}\ell \nu)} = 0.440\pm 0.072\,.   \label{eq:R}
\end{eqnarray}
Both results are consistent with previous measurements performed by
Belle~\cite{Matyja:2007kt} but are larger than the SM values of
$\mathcal R_{\tau/\ell} ^{*,\rm SM} = 0.252(3)$ and $\mathcal
R_{\tau/\ell} ^{\rm SM} = 0.296(16)$ with $3.4\,\sigma$ significance
when the two observables are combined (see
Ref.~\cite{Fajfer:2012jt}). If confirmed these results might point to
NP in (semi)tauonic $b \to c$ transitions.  We accordingly consider a
plausible scenario where the presence of one light color triplet, weak
doublet, scalar leptoquark (LQ) state addresses the aforementioned
discrepancy through a minimal set of couplings to matter. The minimality is
motivated by the fact that the presence of this state was shown to be
too severely constrained by rare decays of charm and strange mesons
and of tau lepton to be able to affect $c \to s \ell^+ \nu$
significantly~\cite{Fajfer:2008tm,Dorsner:2009cu}. It also renders our
analysis more manageable.

The colored scalar we consider appears in various studies of possible extensions
of the SM~\cite{Kronfeld:2008gu,DelNobile:2009st,Dorsner:2009cu,Davidson:2011zn,Barr:2012xb,Arnold:2013cva}. In particular, it can be embedded in $45$-dimensional scalar representation of $SU(5)$~\cite{Georgi:1979df,Giveon:1991zm,Perez:2007rm,Dorsner:2007fy,Dorsner:2009mq} to help provide viable unification of the SM interactions. Inclusion
of additional scalar representations can result in two important
improvements over the original $SU(5)$ setup~\cite{Georgi:1974sy}. Namely, the mass relations between down-type quarks and charged
leptons can be corrected~\cite{Georgi:1979df} and strong, weak and electromagnetic interactions can unify~\cite{Giveon:1991zm,Perez:2007rm,Dorsner:2007fy,Dorsner:2009mq}. We thus investigate whether the minimal set of Yukawa couplings we posit can be made compatible with the mass generation mechanism within a simple $SU(5)$ setup that relies on the use of $45$-dimensional representation.

In this paper we systematically study impact of the colored scalar on
the exclusive $B \to D^{(*)} \tau \bar\nu$ decays, $Z \to b \bar b$
decay, muon anomalous magnetic moment, $\tau$ electric dipole moment
and LFV decays $\mu \to e \gamma$, $\tau \to \mu \gamma, e \gamma$. We
also point out additional observables that can reveal the presence of
this state. These are $t \to c \tau^+ \tau^-$ and $\bar D^0 \to \tau^-
e^+$. We finally discuss viability of having the colored scalar be
part of $45$-dimensional representation of $SU(5)$ and impact of such
an embedding on relevant low-energy phenomenology.

Our work is organized as follows. In Section~\ref{CLASSIFICATION} we
list all color triplet candidates that can have an impact on $b \to c
\tau \bar \nu$ transitions. There we single out the color triplet we
focus on in the rest of our study and specify the minimal set of Yukawa
couplings that is required to address the $B \to D^{(*)} \tau \bar
\nu$ puzzle. The individual LQ contributions to $B \to D \tau \bar
\nu$ and $B \to D^* \tau \bar \nu$ are discussed in
Section~\ref{btoctaunu}, where we present numerical fit to
data. Sections~\ref{Z},~\ref{ems} and~\ref{tautoegamma} are devoted to
the LQ impact on $Z \to b \bar b$, lepton electromagnetic moments and
$\ell \to \ell' \gamma$ decays, respectively. We then investigate the
possibility to have a viable embedding of LQ setup in an $SU(5)$
scenario in Section~\ref{GUT}. The connection between $\ell \to \ell'
\gamma$ and $B \to D^{(*)} \tau \bar \nu$ that follows from the
proposed embedding is discussed in Section~\ref{FEEDBACK}. We proceed to
offer predictions on $B_c \to \tau^+ \nu$ and rare decays in
Section~\ref{PREDICTIONS} and provide conclusions in Section~\ref{CONCLUSIONS}.

\section{Color triplet candidates}
\label{CLASSIFICATION}
The $b \to c \ell \bar \nu$ decay can be mediated, among other
proposals, by color triplet bosons with renormalizable leptoquark
couplings to the SM fermions. These bosons can be either scalars or
vectors with electric charges of $|Q|= 1/3$ and $|Q|=2/3$. We list
quantum numbers of all states with potential contributions to $b \to c
\ell \bar \nu$ decay in Tab.~\ref{tab:states} where we specify their
properties under $SU(3)$ and $SU(2)$ gauge groups as well as
hypercharge $Y$, where hypercharge is defined in terms of electric
charge and weak isospin as $Y = Q-T_3$.  We also show possible scalar
and vector contractions of the SM fermions, omitting the generation
and color indices, and present associated baryon ($B$) and lepton ($L$) numbers,
where applicable.

The vector states are also listed in Tab.~\ref{tab:states} for
completeness but we do not consider them further since it seems
difficult to implement light colored vectors in realistic
scenarios. We also discard two particular scalars---$(3,1)_{-1/3}$ and
$(3,3)_{-1/3}$---since they destabilize
proton~\cite{Dorsner:2012nq}. Of the two remaining scalar states the
one with $Y=1/6$ couples to the right handed neutrino and would not
interfere with the SM amplitudes. This does not pose any
problem since the two observables require enhancement of the
semi-tauonic decays. However, introduction of a light RH neutrino requires
an explanation of its origin and we choose not to pursue this option
in the following. Therefore we dismiss this state as a suitable candidate
for modification of $b\to c\tau\bar\nu$. The only scalar left, on the
other hand, couples to the left-handed neutrino and interferes with
the SM semileptonic amplitude. We denote that scalar with $\Delta
\equiv (3,2)_{7/6}$ in what follows.
\begin{table}[!htcb]
  \centering
  \begin{tabular}{|l|c|p{4cm}|c|c|}
    \hline
$(SU(3),SU(2))_Y$    & spin & \hspace{1cm}LQ couplings & $3B$ & $L$\\\hline    \hline
$(3,2)_{1/6}$ & 0 & $\overline{Q}  \nu_R$, $\overline{d}_R L$ & $+1$
& $-1$\\
$(3,2)_{7/6}$ & 0 & $\overline{Q} \ell_R$, $\overline{u}_R L$ & $+1$ &$-1$ \\
$(3,1)_{-1/3}$ & 0 & $\overline{Q}  i\tau^2 L^C$,  $\overline{d}_R
\nu_R^C$, $\overline u_R \ell_R^C$ & & \\
 $(3,3)_{-1/3}$ & 0 & $\overline Q \tau^i i\tau^2 L^C$ & &\\\hline
$(3,1)_{2/3}$ & 1 & $\overline u_R \gamma_\mu \nu_R$, $\overline Q
\gamma^\mu L$ & $+1$ & $-1$ \\
$( 3,3)_{2/3}$ & 1 & $\overline Q \tau^i \gamma^\mu L$ & $+1$ & $-1$ \\
$(3,2)_{1/6}$ & 1 & $\overline u_R \gamma_\mu i\tau^2 L^C$,
$\overline{Q} \gamma_\mu \nu_R^C$ & $+1$ & $-1$\\
$(\bar 3,2)_{5/6}$ & 1 & $\overline Q \gamma^\mu \ell_R^C$,
$\overline{d_R} i\tau^2 \gamma_\mu  L^C $ & $+1$ & $-1$\\\hline
  \end{tabular}
  \caption{Scalar and vector leptoquarks that trigger $b\to c \ell \bar \nu$ via renormalizable couplings.}
  \label{tab:states}
\end{table}

Yukawa couplings of $\Delta$ to the SM fermions are
\begin{align}
  \label{eq:S76}
  \mc{L} &= \overline \ell_R \,Y\, \Delta^\dagger Q + \bar u_R \,Z\,
  \tilde\Delta^\dagger L + \mrm{H.c.}\,,
\end{align}
where we have used $\tilde \Delta = i \tau_2 \Delta^*$ for the
conjugated state. Transition to the mass basis splits Yukawa couplings of
the weak doublets to two sets of couplings relevant for the upper and
the lower doublet components.  $Y$ and $Z$ in Eq.~\eqref{eq:S76} are Yukawa matrices that differ by a relative
rotation when applied to respective components. $Y$ represents couplings between charged
leptons and down-type quarks while $Z$ connects up-type quarks with charged
leptons. We use the basis where all relative rotations
are assigned to neutrinos and up-type quarks and the transition to such basis is achieved by substituting $\nu_L \to V_\mrm{PMNS}
\nu_L$ and $u_L \to V_\mrm{CKM}^\dagger u_L$, where $V_\mrm{PMNS}$ and $V_\mrm{CKM}$ represent Pontecorvo-Maki-Nakagawa-Sakata (PMNS) and Cabibbo-Kobayashi-Maskawa (CKM) mixing matrices, respectively. It is in this basis that we unambiguously define $Y$ and $Z$. The two components of
colored scalar, i.e., $\Delta^{(2/3)}$ and $\Delta^{(5/3)}$, then couple as
\begin{align}
  \label{eq:L23}
  \mc{L}^{(2/3)}&= (\bar \ell_R Y d_L)\,\Delta^{(2/3)*} + (\bar u_R [Z
  V_\mrm{PMNS}] 
  \nu_L)\,\Delta^{(2/3)}  + \mrm{H.c.}\,,\\
  \mc{L}^{(5/3)}&= (\bar \ell_R [YV_\mrm{CKM}^\dagger]
  u_L)\,\Delta^{(5/3)*} - (\bar u_R Z  \ell_L)\,\Delta^{(5/3)}  + \mrm{H.c.}\,.   \label{eq:L53}
\end{align}
It is non-existence of a diquark bilinear with hypercharge $Y = 7/6$ that makes it
possible to assign lepton and baryon number consistently to $\Delta$.

In the first two generations of quarks and leptons the flavor changing
processes are well fitted with CKM and
PMNS parameters. This agreement is not disturbed by the leptoquarks if we
introduce the minimal set of couplings needed to explain the $b \to c
\tau \bar \nu$ branching fraction. Hence, we only require nonzero coupling of
$\Delta^{(2/3)}$ to
$\bar \tau b$ but not to $\bar b \mu$ or $\bar b e$ as suggested by
nonobservation of anomalies in $b \to c \ell \bar \nu$, with
$\ell=e,\mu$. We also require that only $c$ quark but not $u$ or $t$ couples to neutrinos. These requirements yield
\begin{equation}
  \label{eq:Yuk23}
   Y =
   \begin{pmatrix}
     0 & 0 &0\\
     0 &0&0\\
     0&0& y_{33}
   \end{pmatrix}\,,\qquad
   Z \pmns =
   \begin{pmatrix}
     0&0&0\\
     z_{21}&z_{22}&z_{23}\\
     0&0&0
   \end{pmatrix}\,.
\end{equation}
The $\Delta^{(5/3)}$ Yukawa couplings are related to the 
above ones through CKM and PMNS rotations that induce CKM-suppressed couplings of
$\tau$ to up-type quarks and PMNS-rotated couplings of $c$ quarks to charged leptons. We have
\begin{equation}
\label{eq:Yuk53}
   YV_\mrm{CKM}^\dagger =
   y_{33}\begin{pmatrix}
     0 & 0 &0\\
     0 &0&0\\
     V_{ub}^\ast & V_{cb}^\ast & V_{tb}^\ast
   \end{pmatrix}\,, \qquad
   Z =
   \begin{pmatrix}
     0&0&0\\
     \tilde z_{21}&\tilde z_{22}& \tilde z_{23}\\
     0&0&0
   \end{pmatrix}\,,
\end{equation}
where $\tilde z_{2i}$ are linear combinations of $z_{2j}$ with
$\mc{O}(1)$ coefficients related to the PMNS matrix elements. 
One can, at this stage, regard null-entries in $Y$ and $Z$ to be sufficiently 
small numbers that can thus be neglected in subsequent analyses. We will show later, in Section~\ref{GUT}, that this ansatz is indeed compatible with an $SU(5)$ framework. Also, if the Yukawa couplings at the low energy scale are not too large, the ansatz will be preserved at the high energy scale associated with the scale where gauge couplings unify. This argument, of course, works both ways.

\section{The $(3,2)_{7/6}$ contribution in $b \to c\tau\nu$}
\label{btoctaunu}
The observed anomaly in $R_{\tau/\ell}$ and $R_{\tau/\ell}^{*}$ ratios
has been subject of many
studies~\cite{colangelo,FKN,Crivellin:2012,Datta:2012,damir-nejc-andrey,Choudhury:2012,Pich-Celis:2012,M.Tanaka2012,Ko:2012sv}. Variety
of considered NP scenarios reduces to the effective Lagrangians in
which either new vector/axial-vector and tensor currents, or
(pseudo)scalar density operators are responsible for the measured discrepancy. There are additional observables that could help single out the class
of NP operators preferred by the data~\cite{FKN}. In all these
analyses the effective operator contribution was included into decay
amplitude on individual basis. The model of LQ mediation we
consider here results in scalar/pseudoscalar and tensor contributions
simultaneously. Namely, the relevant effective Hamiltonian for semileptonic $b \to c$ transition
induced by the $(3,2)_{7/6}$ state is
\begin{align}
 \mc{H}^{(2/3)} & = \frac{y_{33} z_{2i}}{2 m_\Delta^2}\,\left[
(\bar  \tau_R \nu_{iL})(\bar c_R b_L) + \frac{1}{4} (\bar  \tau_R
\sigma^{\mu\nu} \nu_{iL})(\bar c_R \sigma_{\mu\nu} b_L) \right],\label{Hamilonianeff}
\end{align}
where $m_\Delta$ is the mass of the LQ component with charge
$|Q|=2/3$ and is also defined as a matching scale for the above
Hamiltonian. (In what follows we assume that $\Delta^{(2/3)}$ and
$\Delta^{(5/3)}$ are degenerate in mass.) This means that the appropriate
Wilson coefficients of scalar and tensor operators, $g_S$ and
$g_T$, are uniquely determined and correlated. The above leptoquark
effective Hamiltonian will affect semileptonic decays with the tau
lepton, but contrary to the SM the final state neutrino is not
necessarily a $\bar\nu_\tau$. The most natural mechanism to enhance $b
\to c \tau \bar\nu$ is to have a constructive interference between the
SM and the LQ amplitudes of $b \to c \tau \bar\nu_\tau$, whereas
pure leptoquark contributions, producing $\bar\nu_e$ and $\bar\nu_\mu$
are negligible. This implies that we employ a Hamiltonian that
includes the SM as well as the LQ contribution to $b \to c
\tau \bar\nu_\tau$ decay
\begin{equation}
  \mc{H}=\frac{4G_F}{\sqrt{2}} V_{cb} \Big[ (\bar\tau_L\gamma^\mu
    \nu_L)(\bar{c}_L\gamma_\mu b_L)+ g_S (\bar{\tau}_R \nu_L)(\bar{c}_R
    b_L) + g_T (\bar{\tau}_R \sigma^{\mu\nu} \nu_L)(\bar{c}_R\sigma_{\mu\nu} b_L) \Big]\,,\label{Hamiltonian}
\end{equation}
where the scalar and tensor effective couplings are related to the
underlying Yukawa couplings at the matching scale
\begin{equation}
  \label{gS}
g_S(m_\Delta)=4g_T(m_\Delta)\equiv\frac{1}{4}\frac{y_{33} z_{23}}{2 m_\Delta^2}\frac{\sqrt{2}}{G_F V_{cb}}.
\end{equation}

Hadronic (pseudo)scalar and tensor operators in Eq.~\eqref{Hamilonianeff} have anomalous dimensions in QCD and dependence of their matrix elements on the renormalization scale is canceled by the scale dependence of the Wilson coefficients at the leading logarithm approximation,
\begin{equation}
\label{eq:rg}
\begin{split}
g_S(m_b)&=\Bigg(\frac{\alpha_S(m_b)}{\alpha_S(m_t)}\Bigg)^{-\frac{\gamma_S}{2\beta_0^{(5)}}}\,\Bigg(\frac{\alpha_S(m_t)}{\alpha_S(m_\Delta)}\Bigg)^{-\frac{\gamma_S}{2\beta_0^{(6)}}}g_S(m_\Delta)\,,\\
g_T(m_b)&=\Bigg(\frac{\alpha_S(m_b)}{\alpha_S(m_t)}\Bigg)^{-\frac{\gamma_T}{2\beta_0^{(5)}}}\,\Bigg(\frac{\alpha_S(m_t)}{\alpha_S(m_\Delta)}\Bigg)^{-\frac{\gamma_T}{2\beta_0^{(6)}}}g_S(m_\Delta)\,.
\end{split}
\end{equation}
Anomalous dimension coefficients are $\gamma_S=-8$, $\gamma_T=8/3$ and
coefficient $\beta_0^{(f)}=11-2/3 n_f$, where $n_f$ is a number of
active quark flavours~\cite{Chetyrkin:1997dh,Gracey:2000am}. The
relation between the Wilson coefficients, given in Eq.~\eqref{gS}, is
valid at matching scale $m_\Delta$ which we set to the reference mass
of $m_\Delta = 500\e{GeV}$. The coefficients are then run to the
beauty quark scale, i.e., $\mu = m_b = 4.2\e{GeV}$, at which the matrix
elements of hadronic currents are calculated. Difference between
running of $g_S$ and $g_T$ modifies the matching scale
relation~\eqref{gS} to
\begin{equation}
  g_T(m_b) \simeq 0.14 \, g_S(m_b)
\end{equation}

\subsection{$B\to D\tau\nu$}
The exclusive decay amplitudes for $B\to D\tau\nu$ transition contain the hadronic matrix element of the vector current, conventionally parametrized by $f_+(q^2)$ and $f_0(q^2)$ form factors
 \begin{equation}
\langle D(p_D) |\bar{c}\gamma^\mu b |\bar{B}(p_B)\rangle=\Big(p_B^\mu+ p_D^\mu-\frac{m_B^2-m_D^2}{q^2}q^\mu\Big)f_+(q
^2)+\frac{m_B^2-m_D^2}{q^2}q^\mu f_0(q^2)\,,
\end{equation}
where $p_{B,D}^\mu$ are four vectors of momenta of $B$ and $D$ mesons and
$q^\mu= p_B^\mu-p_D^\mu$.  The presence of the tensor operator in Eq.~\eqref{Hamiltonian} requires inclusion of an additional form factor $f_T(q^2)$, defined as 
 \begin{equation}
\langle D(p_D)|\bar{c}\sigma^{\mu\nu} b|\bar{B}(p_B)\rangle=-i(p_B^\mu p_D^\nu- p_D^\mu p_B^\nu)\frac{2f_T(q^2)}{m_B+m_D}\,.
\end{equation}
As usual, the scalar matrix element  is related  to $f_0(q^2)$ form factor
$\langle D|\bar{c} b|\bar{B}\rangle=\frac{m_B^2-m_D^2}{m_b-m_c} f_0(q^2)$,  
where $m_b$ and $m_c$ are masses of $b$ and $c$ quarks in
$\overline{MS}$ scheme at the scale
$\mu=m_b$~\cite{damir-nejc-andrey}. The differential branching ratio
can be calculated from the formula
\begin{equation}
\begin{split}
\frac{d\mathcal{B}}{dq^2}(B \to D \ell \bar\nu_\ell)=&
\frac{ \tau_B G_F^2 |V_{cb}|^2 }{192 \pi^3 m_B^3} 
f_+(q^2)^2\Big(1-\frac{m_l^2}{q^2}\Big)^2\lambda^{1/2}\Bigg[\lambda^{2}\Big(1+\frac{m_l^2}{2
  q^2}\Big)\\
&+|g_T|^2\lambda^{2}\frac{2q^2}{(m_B+m_D)^2}\Big(1+\frac{2m_l^2}{q^2}\Big)\left(\frac{f_T(q^2)}{f_+(q^2)}\right)^2\\
&-\lambda^{2}\frac{6m_l}{m_B+m_D} \Re(g_T)\frac{f_T(q^2)}{f_+(q^2)}\\
&+\Big|1-\frac{q^2}{m_l(m_b-m_c)}g_S\Big|^2 \frac{3}{2}\frac{m_l^2}{q^2}(m_B^2-m_D^2)^2\Bigg(\frac{f_0(q^2)}{f_+(q^2)}\Bigg)^2\Bigg]\label{decay-rate-D},
\end{split}
\end{equation}
where $\lambda$ denotes the function $\lambda(m_B^2,m_D^2,q^2) =
(m_B^2-m_D^2-q^2)^2-4m_D^2q^2$. The constant value of the ratio
$f_T(q^2)/f_+(q^2)=1.03(1)$, used in the branching ratio in
Eq.~\eqref{decay-rate-D}, is evaluated in the model
of Ref.~\cite{melikhov-stech}. In the heavy quark limit this ratio is
$f_T(q^2)/f_+(q^2)=1$, as the form factors are equally related to the
Isgur-Wise function,
$f_+(q^2)=f_T(q^2)=\frac{m_B+m_D}{2\sqrt{m_B m_D}}\xi(w)$. We employ the following parametrization of
vector form factors~\cite{falk-neubert1, falk-neubert2, falk-neubert3}
\begin{equation}
\begin{split}
&f_+(q^2)=\frac{G_1(w)}{R_D}\Big|_{w(q^2)},\\
&f_0(q^2)=R_D\frac{1+w}{2}G_1(w)\frac{1+r_D}{1-r_D}\Delta(w)\Big|_{w(q^2)},
\end{split}
\end{equation}
where the constant $R_D$ and the new kinematic variable $w$ are given as
\begin{equation}
\begin{split}
R_D=\frac{2\sqrt{m_B m_D}}{m_B+m_D},\qquad w=\frac{m_B^2+m_D^2-q^2}{2m_B m_D},
\end{split}
\end{equation}
and $r_D$ denotes the ratio of masses of $D$ and $B$ mesons. 
Lattice estimate of the function $\Delta(w)$ is consistent with constant
value $\Delta(w)=0.46\pm 0.02$~\cite{tantalo}. Function $G_1(w)$ can
be found in Appendix~\ref{sec:appBD}.


\subsection{$B \to D^*\tau\bar\nu$}
The decay $B\to D^*\tau\bar{\nu}_\tau$ offers additional tests of the
SM and NP due to the richer spin structure of the final state
particles~\cite{FKN}. Namely, in the case of light lepton in the final
state, one vector and two axial form factors are present in the decay
amplitude, while with $\tau$ in the final state an additional form
factor $A_0(q^2)$ appears. The mediation of the $(3,2)_{7/6}$
leptoquark induces effective Lagrangian containing the tensor operator
that requires knowledge of tensor form factors.  Following notation of
Ref.~\cite{FKN}, the polarization four-vectors of the final state
leptons and $D^*$ vector meson are denoted by $\tilde{\epsilon}_\mu(\lambda)$
and $\epsilon_\mu (\lambda_{D^*})$, respectively. Polarizations take
the following values: $\lambda=0,\pm,t$ and
$\lambda_{D^*}=\pm,0$. Here $t$ denotes the time-like polarization vector.
Standard parametrization of vector and axial hadronic matrix 
elements for the $B \to D^*$ transition is given by
\begin{subequations}
\begin{align}
\langle D^*(p_{D^*},\epsilon)|\bar{c}\gamma_\mu b|{B(p_B)}\rangle &=\frac{2 i V(q^2)}{m_B+m_{D^*}}\;\epsilon_{\mu\nu\alpha\beta}
\epsilon^{*\nu}p_B^\alpha p_{D^*}^\beta \,,
\label{one}\\
\langle D^*(p_{D^*},\epsilon)|\bar{c}\gamma_\mu\gamma_5 b|{B(p_B)}\rangle
&=2m_{D^*}\,A_0(q^2)\frac{\epsilon^*\cdot q}{q^2}q_\mu
+ (m_B+m_{D^*})\,A_1(q^2)\,\left(\epsilon^*_\mu - \frac{\epsilon^*\cdot q}{q^2}q_\mu
\right) \nonumber\\
&\phantom{=}- A_2(q^2)\,\frac{\epsilon^*\cdot q}{m_B+m_{D^*}}\left( (p_B+p_{D^*})_\mu 
-\frac{m_B^2-m_{D^*}^2}{q^2}q_\mu\right)\,.
\label{two}
\end{align}
\end{subequations}
\begin{figure}[t]
\begin{center}
\includegraphics[width=8cm]{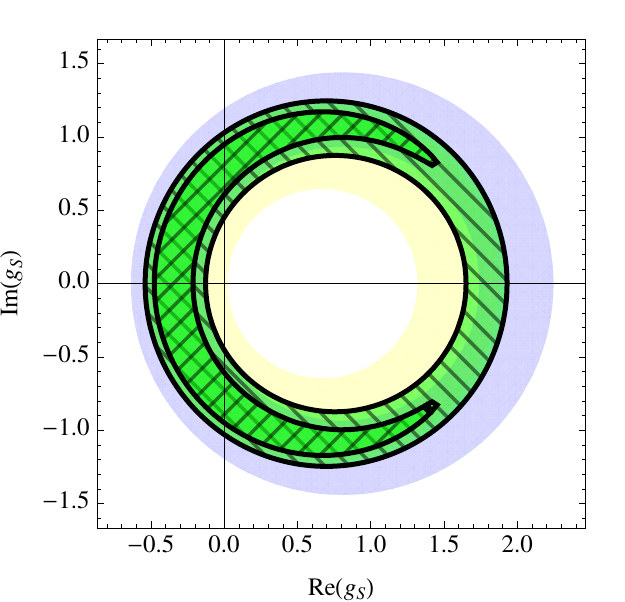}
\end{center}
\caption{Values of the scalar Wilson coefficient $g_S(m_b)$
  ($g_T(m_b)\simeq\,0.14\,g_S(m_b)$) consistent at $2\sigma$ with
  BaBar Collaboration's measurements of ratios $R(D)$ (bright ring) and
  $R(D^*)$ (darker ring).  The 1$\sigma$ (2$\sigma$) region, fitted to
  the two constraints, is doubly (singly) hatched.}
\label{both}
\end{figure}
For the parametrization of the tensor hadronic matrix elements, we adopt the form found in Ref.~\cite{colangelo} that reads
\begin{equation}
\begin{split}
&\langle D^\ast (p_{D^\ast},\epsilon)|\bar{c}\sigma_{\mu\nu}(1-\gamma_5)b|\bar{B}(p_B)\rangle =T_0(q^2)\frac{\epsilon^\ast \cdot q}{(m_B+m_{D^\ast})^2}\epsilon_{\mu\nu\alpha\beta}\,p_B^\alpha\,p_{D^\ast}^\beta+ T_1(q^2)\epsilon_{\mu\nu\alpha\beta} p_B^\alpha \epsilon^{\ast\beta}\\
&+i \Big[T_3(q^2)(\epsilon^\ast_\mu p_{B,\nu}-\epsilon^\ast_\nu p_{B,\mu})+T_4(q^2)(\epsilon_\mu^\ast p_{D^\ast,\nu}-\epsilon_\nu^\ast p_{D^\ast,\mu})+T_5(q^2)\frac{\epsilon^\ast\cdot q}{(m_B+m_{D^\ast})^2}(p_{B,\mu} p_{D^\ast,\nu}-p_{B,\nu} p_{D^\ast,\mu})\Big].\label{tensorialDstar}
\end{split}
\end{equation}
Vector and axial form factors in the heavy quark~(HQ) limit are related to function $h_{A_1}(w)$~\cite{CLN}
\begin{equation}
\begin{split}
A_0(q^2)=\frac{R_0(w)}{R_{D^*}}h_{A_1}(w)|_{w(q^2)}\,,
&\qquad A_1(q^2)=\frac{w+1}{2}R_{D^*}h_{A_1}(w)|_{w(q^2)}\,,\\
A_2(q^2)=\frac{R_2(w)}{R_{D^*}}h_{A_1}(w)|_{w(q^2)}\,,
&\qquad V(q^2)=\frac{R_1(w)}{R_{D^*}}h_{A_1}(w)|_{w(q^2)}\,.
\end{split}
\end{equation}
Functions $R_i(w)$ and $h_{A_1}(w)$ can be found in Appendix~\ref{sec:appBD}. The
tensorial form factors from the parametrization~\eqref{tensorialDstar}
in the HQ limit are related to the
function $h_{A_1}(w)$~\cite{colangelo, isgur-wise}
\begin{equation}
\begin{split}
&T_0(q^2)=T_5(q^2)=0\,,\\
&T_1(q^2)=T_3(q^2)=\sqrt{\frac{m_{D^*}}{m_B}}h_{A_1}(w)|_{w(q^2)}\,,\\
&T_2(q^2)=T_4(q^2)=\sqrt{\frac{m_B}{m_{D^*}}}h_{A_1}(w)|_{w(q^2)}\,.\\
\end{split}
\end{equation}
Using the effective Hamiltonian~\eqref{Hamiltonian}, one finds that
the contributing vector and axial-vector, pseudo-scalar and tensor
amplitudes are given in terms of corresponding hadronic and leptonic
helicity amplitudes~\cite{FKN,colangelo,M.Tanaka2012}:
\begin{equation}
\label{eq:BDstAmp}
\begin{split}
&\mathcal{A}_{V-A}^{\lambda_\tau,\lambda_{D^*}}=\sum_\lambda\,\eta_\lambda H_{V-A,\lambda}^{\lambda_{D^*}}\,L_{V-A,\lambda}^{\lambda_\ell},\\
&\mathcal{A}_{P}^{\lambda_\tau,\lambda_{D^*}}=g_P H_P^{\lambda_{D^*}}L_P^{\lambda_\ell},\\
&\mathcal{A}_{T}^{\lambda_\tau,\lambda_{D^*}}=g_T \sum_{\lambda,\lambda'} \eta_\lambda\eta_\lambda' H_{T,\lambda\lambda'}^{\lambda_{D^*}} L_{T,\lambda\lambda'}^{\lambda_l}.
\end{split}
\end{equation}
The metric factor $\eta$ has values $\eta_{\pm,0}=1$ and $\eta_t=-1$ \cite{martin-wade}.
Hadronic and leptonic helicity amplitudes are defined in Appendix~\ref{sec:appBD}.

The branching ratio is calculated after integrating the following
formula over $q^2$ and angle $\theta_l$~\cite{martin-wade}:
\begin{equation}
d\mathcal{B} =\frac{\tau_B G_F^2 |V_{cb}|^2}{(8\pi)^3\, m_B^2} | \mathbf{p}_{D^*}| \Bigg(1-\frac{m_l^2}{q^2}\Bigg)\sum_{\lambda_\tau,\lambda_{D^*}}\Big|\mathcal{A}^{\lambda_\tau,\lambda_{D^*}}\Big|^2 dq^2\,d\cos\theta_l.
\end{equation}
We constrain the allowed values of tensor and scalar Wilson's
coefficients using BaBar's measurements of the ratios
$\mc{R}_{\tau/\ell}=\mathcal{B}(B\to D\tau\nu)/\mathcal{B}(B\to D\ell\nu)$
($\mc{R}^\ast_{\tau/\ell}=\mathcal{B}(B\to D^\ast\tau\nu)/\mathcal{B}(B\to
D^\ast\ell\nu)$) as shown in
Fig.~\ref{both}, where also the result of the fit to both ratios is shown. We derive $1\sigma$ range for the Wilson
coefficient $g_S$ at the low scale
\begin{equation}
 \label{eq:gSrange}
  g_S(m_b) = -0.37^{+0.10}_{-0.07}\,,
\end{equation}
where we have assumed $g_S$ to be real in estimating the error bars. The
coupling $g_S$ at the matching scale, defined in Eq.~\eqref{gS}, is
rescaled by factor $0.64$ with respect to the above value due to QCD corrections~\eqref{eq:rg}.

\section{$Z \to b \bar b$}
\label{Z}
\begin{figure}[!htbp]
  \centering
  \includegraphics[width=0.6\textwidth]{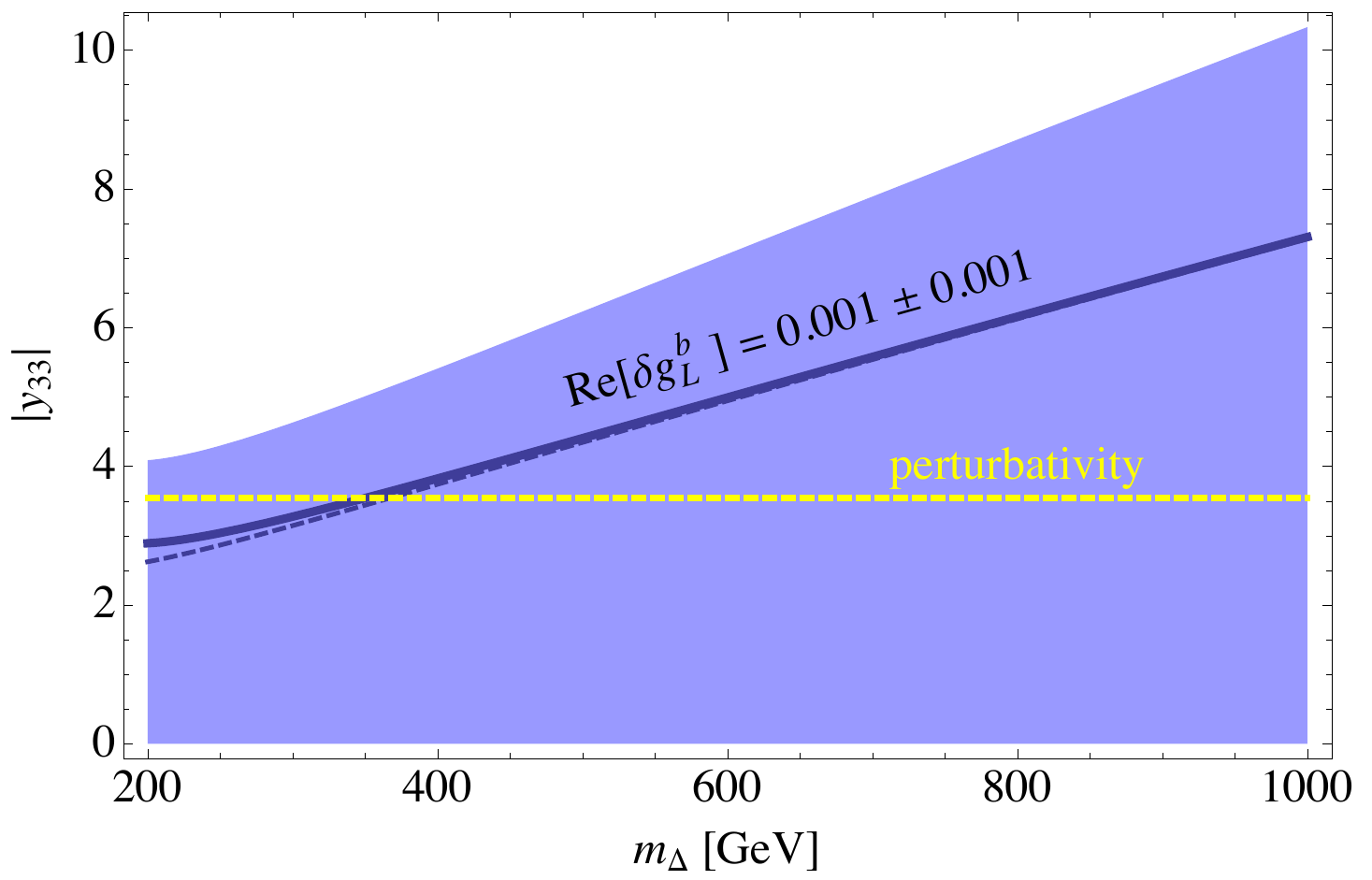}
  \caption{Shaded region in the $m_\Delta$--$|y_{33}|$
    plane reproduces the $1\,\sigma$ interval for $\re[\delta g_L^b] = 0.001 \pm
    0.001$. Thick solid line denotes the central value while accompanying dashed
    line represents the asymptotic form. Bright horizontal line is a
    bound above which coupling $|y_{33}|$ is nonperturbative at
    scale $m_\Delta$.}
  \label{fig:Zbb}
\end{figure}

The LEP experiment measured precisely decay modes of $Z$ bosons to $f
\bar f$ pairs. Standard parameterization of the $Zb \bar b$ renormalizable
coupling is adopted
\begin{equation}
\label{eq:Zbbdef}
\mc{L}_{Zb\bar b} = \frac{g}{c_W} Z^\mu 
    \bar b \gamma_\mu \left[(g_L^b + \delta g_L^b) P_L +  (g_R^b +
      \delta g_R^b) P_R \right] b\,.
\end{equation}
$SU(2)$ coupling is denoted by $g$, $c_W$ is the cosine of the
Weinberg angle and $P_{L,R} = (1 \pm \gamma_5)/2$ are the chiral
projectors. At the SM tree-level, the couplings are $g_L^{b0} =
-1/2+s_W^2/3$ and $g_R^{b0} = s_W^2/3$. Higher-order electroweak
corrections that are contained within $g_{L,R}^b$ get largest
contributions from top quark in loops. A recent electroweak fit
that includes updated theoretical predictions and new results from LHC
points to tensions in the $Z \to b\bar b$ observables reaching above
$2\,\sigma$ significance in $R_b$ and
$A_\mrm{FB}^b$~\cite{Batell:2012ca}~(see
also~\cite{Fajfer:2013wca,Freitas:2012sy,Eberhardt:2012gv,Baak:2012kk,Aad:2012tfa,Chatrchyan:2012ufa}).
The shifts with respect to the SM values of couplings are found to be
\begin{align}
  \label{eq:dgb}
  \delta g_L^b = 0.001 \pm 0.001 \,,\qquad \delta g_R^b = (0.016 \pm
  0.005) \cup (-0.17 \pm 0.005)\,.
\end{align}
$\Delta^{(2/3)}$ component possesses a
possibly large coupling $y_{33}$ to $b \tau$ pair that contributes to $Z
\to b \bar b$ amplitude at order $|y_{33}|^2$ and thus allows one to constrain it directly. The LQ correction to the left-handed coupling is
\begin{align}
  \label{eq:dgLb}
  \delta g_L^b(y_{33}) = \frac{|y_{33}|^2}{384 \pi^2} \left[ g_0(x) + s_W^2 g_2(x) \right]\,,
\end{align}
where $x = m_\Delta^2/m_Z^2$, and we take $m_Z = 91.2$\,GeV and $s_W^2
= 0.231$. Asymptotic expansion of functions $g_0$ and $g_2$ at large $x$ is
\begin{align}
  g_0(x) &\simeq -\frac{2}{3x}\,,\nn\\
   g_2(x) &\simeq \frac{8}{x} (\log x+2/9+i \pi )\,,\nn
\end{align}
while their full expressions can be found in
Appendix~\ref{sec:Zbbapp}. Although $\delta g_L^b$ develops an
imaginary part due to on-shell $\tau$ leptons, the constraint from the
electroweak fit~\eqref{eq:dgb} is sensitive to the interference term
between the approximately real $g_L^b$ and complex $\delta
g_L^b(y_{33})$, and therefore only the real part of $\delta
g_L^b(y_{33})$ enters the prediction. The constraint 
\begin{equation}
  \re[\delta g_L^b(y_{33})] = 0.001 \pm 0.001\,,
\end{equation}
is shown in the $m_\Delta$--$|y_{33}|$ plane on Fig.~\ref{fig:Zbb}.
In the range of $m_\Delta$ shown there we approximate
the central value (thick line) by a linear function that is accurate to within $5\,\%$:
\begin{equation}
 |y_{33}|_\mrm{central} = 1.57 + 2.86 \frac{m_\Delta}{500\e{GeV}}\,.
\end{equation}
For $m_\Delta$ above $300\e{GeV}$ large portion of preferred
$|y_{33}|$ range lies within the nonperturbative regime that is situated above a
bright dashed line in Fig.~\ref{fig:Zbb}. In order to maintain a predictable
setup we assume that coupling $y_{33}$ is perturbative, i.e., $|y_{33}| <
\sqrt{4\pi}$. Contributions to $\delta g_R^b$ in this scenario are
further suppressed by $m_b^2/m_\Delta^2$ and can be safely neglected. Later on,
when we discuss the grand unified theory (GUT) embedding, we will find that the perturbativity of Yukawa couplings all the way to the scale of unification puts more stringent upper bound on $|y_{33}|$ at the low energy scale.

On the other hand, corrections to the self-energies of electroweak
gauge bosons that are measured by the oblique parameters $S$, $T$, and
$U$ do not enforce relevant constraints on the LQ Yukawa 
couplings, as long as the two mass eigenstates are approximately
degenerate~\cite{Davidson:2010uu}.

We have also checked what further information on the coupling
  $y_{33}$ could be extracted from atomic parity violation
  experiments. In our model the ansatz for couplings ensures there are
  no tree-level contribution to parity-odd dimension-6 operators with
  flavor $\bar u u e^+ e^-$ that can be generated with the leptoquark
  only through loop-induced $Z\bar u u$ vertex.  One can see from
  Eq.~\eqref{eq:Yuk53} that loop diagrams with $\Delta^{(5/3)}$ and
  $\tau$ leading to $Z \to \bar u u$ are suppressed by $|V_{ub}|^2$ and it is due to this
  suppression that the constraints on $Z\bar u u$ vertex, presented
  in~\cite{Gresham:2012wc}, play no relevant role in constraining the
  considered model.

\section{Lepton electromagnetic moments}
\label{ems}
Three form factors encompass the structure of $\ell\ell \gamma$ vertex
and generalize the tree-level QED vertex~(see, for example, Eq.~(17) of
Ref.~\cite{Jegerlehner:2009ry} and Eq.~(2.2)
in Ref.~\cite{Pospelov:2005pr})
\begin{equation}
  \label{eq:GammaMu}
 -ie\,\bar u_\ell(p+q) \gamma^\mu u_\ell(p)\to
-ie\,\bar u_\ell(p+q)  \left[F_E(q^2) \gamma^\mu + \frac{F_M^\ell(q^2)}{2m_\ell}
    i\sigma^{\mu\nu} q_\nu + F_d^\ell(q^2) \,\sigma^{\mu\nu} q_\nu \gamma_5\right] u_\ell(p)\,.
\end{equation}
The lepton vertex with electromagnetic field will be modified by penguin diagrams
involving virtual exchanges of $\Delta^{(5/3)}$ and charm quark. For the muon the $Z$ couplings in
Lagrangian~(Eqs.~\eqref{eq:L53} and \eqref{eq:Yuk53})
will contribute to the magnetic moment $F_M(q^2)$, but not to the
electric dipole moment $F_d^\ell(q^2)$ which is a CP violating quantity and
requires two different couplings with different complex phases. For the
$\tau$, however, there are two distinct couplings---$YV_\mrm{CKM}$ and
$Z$---at our disposal that
are sufficient to generate the electric dipole moment~(EDM).

\subsection{Muon $(g-2)$}
\begin{figure}[!htbp]
  \centering
  \includegraphics[width=0.6\textwidth]{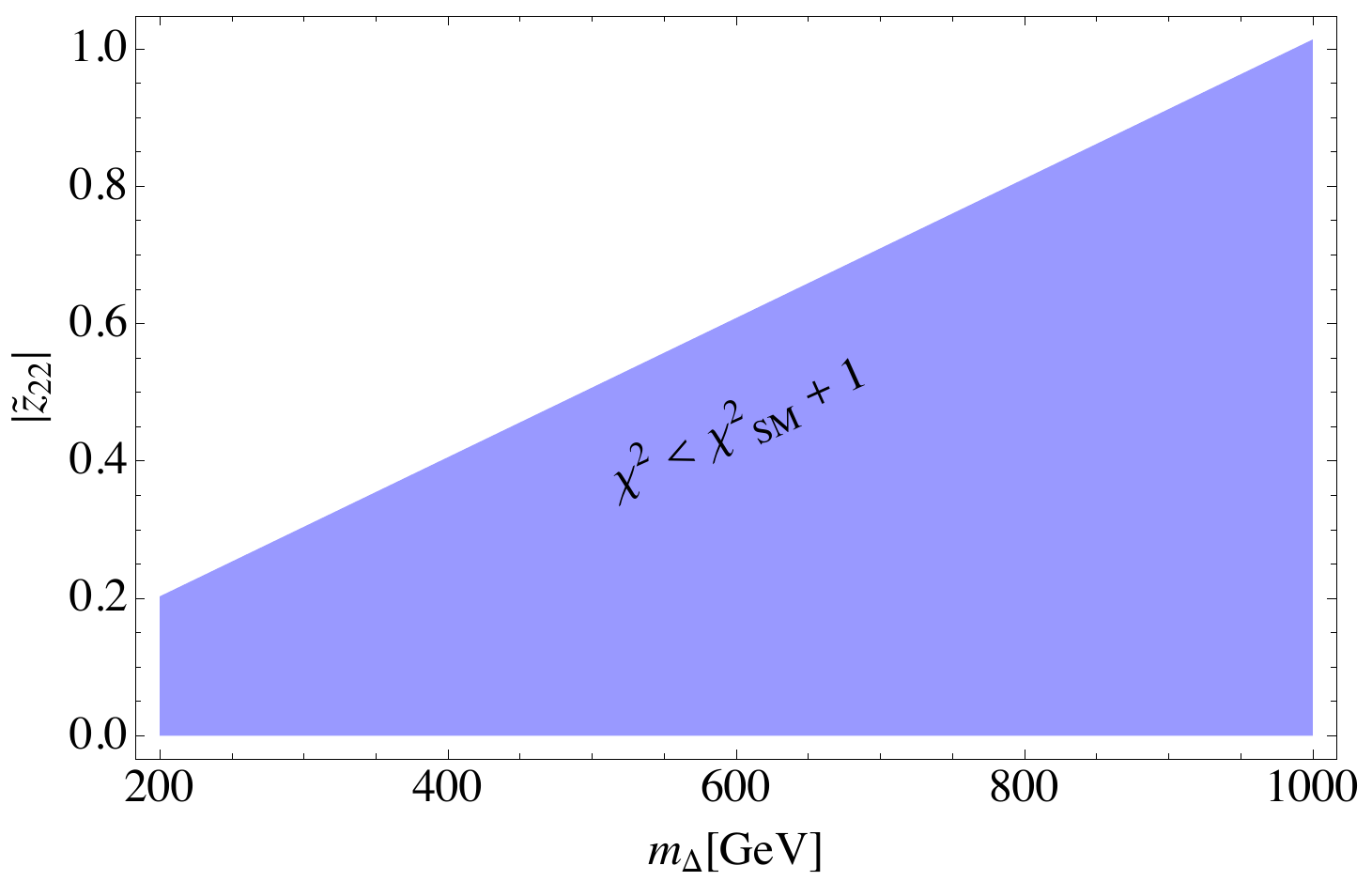}
  \caption{Allowed range of $|\tilde z_{22}|$ by the experimental value
    of the muon magnetic moment (shaded region).}
  \label{fig:gmu}
\end{figure}

The two penguin diagrams and field renormalization factors are
calculated employing the Feynman rules listed in
Tab.~\ref{tab:rules}.  We decompose the amplitude as in
Eq.~\eqref{eq:GammaMu} and identify $F_M(q^2)$. Then we set $q^2 = 0$
and expand $F_M(0)$ to first order in $m_\mu^2$. The muon anomalous
moment gets shifted due to new contribution of $\Delta^{(5/3)}$ by
\begin{equation}
  \label{eq:amu1}
  \delta a_\mu  \equiv F_M^\mu(q^2 = 0) = -\frac{N_c |\tilde z_{22}|^2 m_\mu^2}{16\pi^2 m_\Delta^2}
  \left[Q_c F_q(x) + Q_\Delta F_{\Delta}(x)\right]\,,
\end{equation}
where $x = m_c^2/m_\Delta^2$ and $N_c=3$. Functions $F_q(x)$ and $F_\Delta(x)$ are
\begin{align}
  F_q(x) &= \frac{x^3-6 x^2+3 x+6 x \log x+2}{6
    (x-1)^4}\,,\nn\\
  F_\Delta(x) &= \frac{ 2 x^3+3 x^2-6 x^2 \log x-6 x+1}{6
    (x-1)^4}\,.\nn
\end{align}
The above formulas agree with Ref.~\cite{Cheung:2001ip}. Comparing
the prediction of $\delta a_\mu$ with the value reported by the PDG~\cite{Beringer:1900zz},
\begin{equation}
  \delta a_\mu^\mrm{exp-SM} = (287 \pm 80)\E{-11}\,,
\end{equation}
we observe that $a_\mu$ is pulled further away from the experimental
value. The best fit point corresponds to the SM limit, and is
$3.6\,\sigma$ below the measured value ($\chi^2_\mrm{SM} =
12.87$). We determine the allowed $1\,\sigma$ range from the
condition $\chi^2 - \chi^2_\mrm{SM} \leq 1$ that translates to a
constraint
\begin{equation}
  |\delta a_\mu| < 10.9\E{-11}\,,
\end{equation}
or put in terms of $\tilde z_{22}$ (see Fig.~\ref{fig:gmu})
\begin{equation}
  \label{eq:gmuconstr}
  |\tilde z_{22}| < 0.51\, \frac{m_\Delta}{500\e{GeV}}\,.
\end{equation}
This bound will also need to be reconsidered in view of the GUT embedding we present in Section~\ref{GUT}. It can, however, be considered as a correct upper limit if one discusses simple extension of the SM with additional LQ with relevant couplings given in Eqs.~\eqref{eq:S76} and~\eqref{eq:Yuk23}.

\subsection{$\tau$ electric dipole moment}
Penguin diagrams where $\Delta^{(5/3)}$ scalar couples to fermions with
different couplings may generate EDM. In this section we focus
exclusively on contributions proportional to product of $Y V_{CKM}^\dagger$ and
$Z$. The two couplings have opposite chiralities and at least one
helicity flip, proportional to mass, is needed on the internal charm
propagator.  As a result both penguin diagrams are finite and
result in the following expression for the $\tau$ EDM
\begin{equation}
  \label{eq:tauEDM}
  d_\tau \equiv e F_d^\tau(q^2 = 0) = e\, \frac{m_c \im\left[V_{cb}
      y_{33}^* \tilde{z}_{23}^*\right|}{32\pi^2 m_\Delta^2} \left[ 1 + 4
   \log \frac{m_c^2}{m_\Delta^2}\right]\,.
\end{equation}
At the moment, the best bounds from Belle experiment are orders of
magnitude too weak to directly probe the parameter range, preferred by
$B \to D^{(*)} \tau \bar\nu$~\cite{Inami:2002ah}. From the $90\,\%$
confidence level range $-2.2 \E{-17}e\e{cm} < d_\tau < 4.5
\E{-17}e\e{cm}$ one finds only a weak bound $ -1.5\E{3} <
\im\left[V_{cb} \tilde z_{23}^* y_{33}^*\right] < 3.0\E{3} $, valid at a typical
mass $m_\Delta = 500\e{GeV}$. Of course, couplings of these magnitudes are
meaningless. However, we can turn around the reasoning and
require that all of them stay perturbative. Then we find from
$|\tilde z_{23} y_{33}| < 4\pi$ that the upper bound on tau EDM in
the perturbative setting is $|d_\tau| < 2.6\E{-21}\,e\e{cm}$.

\section{$\ell \to \ell' \gamma$ decays}
\label{tautoegamma}
The decay $\tau \to \ell \gamma$ is mediated by a loop diagram with
$\Delta^{(5/3)}$ scalar and a charm quark. The electromagnetic dipole
transition amplitude can be written as a sum over two photon
polarization amplitudes
\begin{equation}
  \label{eq:tau-ell-gamma}
 \mc{A}_{\tau \to \ell \gamma} = \bar \ell (p')
 \sigma^{\mu\nu}\,\epsilon_\mu^*(q) q_\nu \left( A_{\tau\ell} P_R + B_{\tau\ell}
   P_L \right)
   \tau(p)\,,
\end{equation}
with $q=p-p'$ and $\epsilon$ the polarization of the photon.
Polarization-averaged branching ratio is then
\begin{equation}
  \mc{B}(\tau \to \ell \gamma) = \frac{\tau_\tau}{16\pi}
  \frac{(m_\tau^2-m_\ell^2)^3}{m_\tau^3} \left(|A_{\tau\ell}|^2 + |B_{\tau\ell}|^2\right)\,.\nn
\end{equation}
Expressed in terms of the underlying couplings and masses, and to
leading order in $m_c$ and $m_\tau$, the polarization amplitudes are
\begin{align}
A_{\tau \ell} &= \frac{-N_c e }{48\pi^2 m_\Delta^2}  \left[ m_c\,V_{cb} y_{33}^*
   \tilde{z}_{2\ell}^* (1 + 4
   \log x_c ) + \frac{m_\tau}{2}\tilde{z}_{23}
   \tilde{z}_{2\ell}^* ( 3 + 4 x_c
   \log x_c ) \right] \,,\label{eq:Al}\\
B_{\tau\ell} &= 0\,,
\end{align}
where $x_c = m_c^2/m_\Delta^2$. To connect the two vertices with
opposite chiralities of quarks (couplings $V_{cb} y_{33}$ and $\tilde
z_{2\ell}$) a helicity flip on the charm quark propagator is
required. If one considers a contribution proportional to $\tilde
z_{2\ell} \tilde z_{23}$ we need a helicity flip on the lepton legs,
where the $\tau$ mass insertion contributes to $A_{\tau\ell}$ while the
contribution of $\mu$ mass insertion to $B_{\tau\ell}$ is negligible. The decay $\mu \to e \gamma$ proceeds exclusively through the
second term in~\eqref{eq:Al} and is relatively suppressed by a factor
$m_\mu/m_\tau$
\begin{align}
A_{\mu e} &= \frac{-N_c e m_\mu }{96\pi^2 m_\Delta^2} \,\tilde{z}_{22}
   \tilde{z}_{21}^* ( 3 + 4 x_c \log x_c ) \,,
\end{align}
whereas $B_{\mu e}$ is rendered negligible due to helicity flip
on the electron leg.

The best experimental bounds on LFV radiative decays of $\tau$ were
presented by BaBar collaboration in Ref.~\cite{Aubert:2009ag}. At 90\,\%
confidence level they read
\begin{equation}
  \label{eq:tau-ell-gammaEXP}
   \mc{B}(\tau \to e \gamma) < 3.3\E{-8}\,,\quad \mc{B}(\tau \to \mu \gamma) < 4.4\E{-9}\,,
\end{equation}
and severely constrain the combination of couplings, present in
Eq.~\eqref{eq:Al}. The experimental upper limits for $\mu \to e
\gamma$ branching ratio are orders of magnitude more stringent and in
conjuction with the small width of the muon they compensate the for
the $m_\mu$ suppression in sensitivity. We rely on the latest result
from the MEG experiment obtained from data collected in years
2009--2011~\cite{Adam:2013mnn},
\begin{equation}
  \label{eq:me-e-gammaEXP}
   \mc{B}(\mu \to e \gamma) < 5.7\E{-13}\,,\qquad \textrm{at 90 \% C.L.}\,.
\end{equation}
These constraints and our Yukawa ansatz will both be interpreted in the GUT framework we introduce next.

\section{$(3,2)_{7/6}$ in GUT framework}
\label{GUT}
The most natural setting for the scalar leptoquark $\Delta \equiv (3,2)_{7/6}$ is within a framework of matter unification~\cite{Pati:1974yy,Georgi:1974sy}. For example, if we resort to a language of $SU(5)$ we can show that $\Delta$ with couplings to the SM fermions can be found in $45$- and $50$-dimensional representations of that group~\cite{Slansky:1981yr}. However, only $45$-dimensional representation can simultaneously generate both types of couplings in Eq.~\eqref{eq:S76}. If, on the other hand, $\Delta$ originates solely from $50$-dimensional representation we could reproduce only those couplings of $\Delta$ to matter that are proportional to the $Y$ matrix entries. It is the former scenario that we will thoroughly study later on. 

One might prefer to discuss the origin of $\Delta$ within an $SO(10)$ framework. In the $SO(10)$ setup the relevant representations that couple to matter and contain $\Delta$ are $120$- or $126$-dimensional ones. In particular, there is one scalar found in both $120$- and $126$-dimensional representations of $SO(10)$ that couples to the matter as if it is $\Delta$ from $45$-dimensional representation of $SU(5)$. $\Delta$ from $50$-dimensional representation of $SU(5)$, on the other hand, can only be embedded into $126$-dimensional representation of $SO(10)$. Either way, the $SO(10)$ origin of $\Delta$ would be imprinted on its couplings to matter in a flavor basis on top of the $SU(5)$-like behavior. Namely, it is well-known that $120$- and $126$-dimensional representations of $SO(10)$ couple to the matter antisymmetrically and symmetrically, respectively. The $\Delta$ couplings to matter would simply inherit these properties.

Let us now look in detail whether a particular low-energy ansatz with Yukawa couplings, as given in Eq.~\eqref{eq:S76}, is compatible with the idea of grand unification. In what follows we exclusively use the language of $SU(5)$ to specify relevant operators. In view of the comments put forth in the previous paragraph it is easy to see that a switch to the $SO(10)$ framework is straightforward, if and when needed. 

We take $\Delta$ to originate from $45$-dimensional scalar representation. Again, this is the only option available to generate the LQ interactions proportional to the $Z$ coupling matrix. The same scalar representation is actually needed to generate viable fermion masses. It is this fact that allows one to relate the LQ couplings to fermion masses as we show later in this section. Realistic charged fermion masses require presence of an additional $5$-dimensional scalar representation~\cite{Georgi:1979df}. We accordingly assume presence of both scalar representations and refer to them as $\mathbf{5}$ and $\mathbf{45}$ using their dimensionality. We furthermore denoted vacuum expectation values (VEVs) of neutral components of Higgs doublets in $\mathbf{5} (\equiv \mathbf{5}^\alpha)$ and $\mathbf{45} (\equiv \mathbf{45}^{\alpha \beta}_\gamma)$ with $v_5$ and $v_{45}$, respectively. Here, $\alpha, \beta, \gamma(=1,\ldots,5)$ represent $SU(5)$ indices. Of course, to address observed masses of charged leptons and down-type quarks both VEVs are needed. Our normalization is such that $|v_5|^2/2+12|v_{45}|^2=v^2$, where $v(=246\,\mathrm{GeV})$ is the electroweak VEV. In other words, we take $\langle\bm{5}^5\rangle= v_5 / \sqrt{2}$ and $\langle\bm{45}^{15}_{1}\rangle= \langle\bm{45}^{2 5}_{2}\rangle=\langle\bm{45}^{3 5}_{3}\rangle = v_{45}/\sqrt{2}$~\cite{Dorsner:2011ai}. 

We turn our attention to the $\Delta$ couplings to matter that are proportional to $Z$. The $SU(5)$ contractions relevant for these Yukawa couplings are 
\begin{align}
\label{eq:111}
&(Y_1)_{ij} \mathbf{10}_i \overline{\mathbf{5}}_j \mathbf{45},\\
\label{eq:112}
&(Y_3)_{ij} \mathbf{10}_i \overline{\mathbf{5}}_j \mathbf{5},
\end{align}
where $\mathbf{10}_i$ and $\overline{\mathbf{5}}_i$ together comprise an entire generation of fermions~\cite{Georgi:1974sy}. $Y_1$ and $Y_3$ are, at this stage, arbitrary $3 \times 3$ matrices in flavor space with $i,j(=1,2,3)$ being corresponding generation indices.

We denote unitary transformations of the down-type quark fields to be $D_L$ and $D_R$, where subscripts $L$ and $R$, from now on, correspond to redefinitions of the left- and right-handed fields, respectively. These rotations take the down-type quark fields from a flavor into a mass eigenstate basis. For the up-type quark (charged lepton) sector we similarly adopt $U_L$ and $U_R$ ($E_L$ and $E_R$) to be appropriate unitary matrices. We further assume that neutrinos are Majorana particles and accordingly denote unitary matrix that defines the neutrino mass eigenstates with $N$. Exact mechanism of the neutrino mass generation is not important for our discussion.

The low-energy ansatz we want to implement in the $SU(5)$ framework is completely specified through the following set of transformations: $\nu_L \to V_\mrm{PMNS} \nu_L$ and $u_L \to V_\mrm{CKM}^\dagger u_L$. This implies that the unitary rotations of the left-handed down-type quarks ($D_L$) and charged leptons ($E_L$) are both diagonal unitary matrices at the $SU(5)$ symmetry breaking scale. The same ansatz also fixes $U_L$ and $N$ in terms of low-energy observables $V_\mrm{CKM}^\dag$ and $V_\mrm{PMNS}$, respectively. The entries of $V_\mrm{CKM}^\dag$ and $V_\mrm{PMNS}$ are, on the other hand, well-measured observables. What is not {\it a priori} known, however, are rotations in the right-handed fermion sector, i.e., $U_R$, $D_R$ and $E_R$. We demonstrate in what follows that all angles in $U_R$, $D_R$ and $E_R$ are specified through the ansatz of Eq.~\eqref{eq:Yuk23}, except for one angle in $U_R$ matrix, within the proposed $SU(5)$ framework. This behavior can be traced back to a restrictive form of $Z$ couplings. Moreover, we show that the entries of the $Z$ coupling matrix exhibit hierarchy that is similar to the mass hierarchy present in the charged lepton and the down-type quark sectors. To simplify our discussion, we take both these sectors to be real. This we do because low-energy phenomenology requires no phases whatsoever.

What we find, after we impose the ansatz of Eq.~\eqref{eq:S76} on contractions in Eqs.~\eqref{eq:111} and~\eqref{eq:112} at the GUT scale, are the following two relations that connect fermion mass matrices of down-type quarks and charged leptons with the original Yukawa couplings:  
\begin{align}
2 M_D^\mathrm{diag} D_R^T&=- 2 Y_1 v_{45}- Y_3 v_{5}\,,\\
2 E_R M_E^\mathrm{diag}&=6 Y_1 v_{45}- Y_3 v_{5}\,.
\end{align}
Here, $M_D^\mathrm{diag}$ ($M_E^\mathrm{diag}$) is a diagonal mass matrix for down-type quarks (charged leptons) and we take both $v_{5}$ and $v_{45}$ to be real. Note that the relations in question contain only the right-handed unitary transformations $D_R$ and $E_R$. We proceed by identifying a connection between $Y_1$ and $Z$ to be $Y_1=-U_R Z$. This, then, leads us to the following matrix equation
\begin{equation}
\label{eq:1}
M_D^\mathrm{diag} D_R^T-E_R M_E^\mathrm{diag}=4 U_R Z v_{45}\,.
\end{equation}
If $Z$ that at the GUT scale has a form given in Eq.~\eqref{eq:Yuk53} is to satisfy this set of equations our ansatz would be compatible with the idea of $SU(5)$ grand unification.

Let us count the number of relevant parameters in Eq.~\eqref{eq:1}. Clearly, since $Z$ has only one row of elements different from zero we have a situation where elements of only one column of $U_R$ enter the matrix equation. The entries of that column can be parametrized with two angles we denote with $\phi$ and $\theta$. More specifically, we take $(U_R)_{21}=\sin \theta \cos \phi$, $(U_R)_{22}=\sin \theta \sin \phi$ and $(U_R)_{23}=\cos \theta$. As for $D_R$ and $E_R$ matrices, each one contains three angles. We accordingly parametrize them via angles $\theta^D_i$ and $\theta^E_i$, $i=1,2,3$, where we define generic orthogonal matrix $O(\theta_1,\theta_2,\theta_3)=O_3(\theta_3)\,O_2(\theta_2)\,O_1(\theta_1)$ through
\begin{equation}
O_3(\theta)=\left( \begin{array}{ccc}
1   & 0   & 0 \\
0 & \cos \theta & \sin \theta\\
0 & -\sin \theta & \cos \theta
\end{array} \right),\,  
O_2(\theta)=\left( \begin{array}{ccc}
\cos \theta   & 0   & \sin \theta \\
0 & 1 & 0\\
-\sin \theta & 0 & \cos \theta
\end{array} \right),\,  
O_1(\theta)=\left( \begin{array}{ccc}
\cos \theta   & \sin \theta & 0 \\
-\sin \theta & \cos \theta & 0\\
0 & 0 & 1
\end{array} \right).
\end{equation}
Of course, the elements of $Z$ matrix, i.e., $\tilde z_{21}$, $\tilde z_{22}$ and $\tilde z_{23}$, are not fixed and we consider them to be free parameters as well. Note that they scale with a common factor $v_{45}$. If $\Delta$ is a mixture of states that reside in both $45$- and $50$-dimensional representations the common factor would then also contain relevant parameter that describes a level of that mixing. All in all, the number of real parameters is eleven whereas the number of equations is nine. 

Eq.~\eqref{eq:1} should be satisfied at the scale of $SU(5)$ unification. This scale depends on the details of the underlying model. To provide relevant input parameters we resort to a particular $SU(5)$ scenario~\cite{Perez:2007rm} that besides $\mathbf{10}_i$, $\overline{\mathbf{5}}_i$, $\mathbf{45}$ and $\mathbf{5}$ encompasses two $24$-dimensional representations. One of these is a scalar representation needed to break  $SU(5)$ symmetry in the usual manner. The other one is fermionic in nature and is required to generate realistic masses and mixing angles in the neutrino sector~\cite{Bajc:2006ia}. Unification within this setup has been extensively studied in Refs.~\cite{Perez:2008ry,Dorsner:2009mq}. Our numerical input for the masses of down-type quarks and charged leptons at the scale of unification reads $m_d = 0.00105$\,GeV, $m_s = 0.0187$\,GeV, $m_b = 0.782$\,GeV, $m_e = 0.000435$\,GeV, $m_\mu = 0.0918$\,GeV and $m_\tau = 1.56$\,GeV. (See Tab.~VIII of Ref.~\cite{Dorsner:2011ai} for details on how these input parameters are generated.) We stress that these masses do not vary significantly with respect to small changes in the scale of unification within this scenario. In fact, we obtain similar values for the model where fermionic $24$-dimensional representation is replaced with $15$-dimensional scalar representation~\cite{Dorsner:2007fy}. 

What we find numerically is that there exists only one satisfactory solution to Eq.~\eqref{eq:1} that reads $v_{45} \tilde z_{21}=0.012\e{GeV}$, $v_{45} \tilde z_{22}=0.16\e{GeV}$ and $v_{45} \tilde z_{23}=0.50\e{GeV}$. This solution corresponds to the following values of angles: $\theta^D_1=2.807$, $\theta^D_2=0.062$, $\theta^D_3=0.942$, $\theta^E_1=0.198$, $\theta^E_2=0.038$, $\theta^E_3=-2.988$, $\theta=0.129$ and $\phi=-1.342$. The very existence of this fit implies that the simplest mechanism to generate viable fermion masses in the down-type quark and the charged lepton sectors in $SU(5)$ is compatible with our ansatz. Moreover, the ansatz yields ratios between $\tilde z_{21}$, $\tilde z_{22}$ and $\tilde z_{23}$ that mimic mass hierarchy in the down-type quark and the charged lepton sectors with $\tilde z_{23}$ being a dominant element. Namely, we find that 
\begin{equation}
  \label{eq:zhier}
  \tilde z_{21} \,:\,  \tilde z_{22} \,:\,  \tilde z_{23} = 0.024 \,:\, 0.32  \,:\, 1\,.
\end{equation}
We stress that these results are completely independent from the up-type quark and the neutrino sectors, where CKM and PMNS mixing parameters reside, respectively.  

The fact that the numerical solution is unique even though available parameters outnumber equations can be traced back to the hierarchical property of the down-type quark and the charged lepton sectors. Namely, to address substantial mass splittings for three generations one needs at least three non-zero entires in the $Z$ matrix. We could thus argue that our ansatz belongs to a class of the most minimal ones that are still compatible with the $SU(5)$ unification.

Note that we did neglect the influence of running on null entries in matrix $Z$. The main reason we do that is because we do not {\it a priori} know the strength of the Yukawa coupling entries in $Z$ at the low energy scale. The running of Yukawa couplings and subsequent numerical fit would require iterative approach within very precisely defined model that is beyond the scope of this work. One should accordingly view our numerical analysis as a first approximation of the full-fledged model dependent study. However, we do find that for the values of $|\tilde z_{2i}|\lesssim0.5$, $i=1,2,3$, at the low energy scale we preserve the perturbativity of Yukawa couplings all the way to the GUT scale while preserving the ansatz given in Eq.~\eqref{eq:Yuk23}. Self-consistency of our analysis would thus require that we consider regime in which $v_{45} > \mathcal{O}(1)$\,GeV. The same study yields that $|y_{33}|$, at the low energy scale, should be below $0.8$ in order to preserve perturbativity. In fact, in the regime we advocate, the radiative corrections on null entries in $Z$ can indeed be neglected. Note that the estimate of the radiative correction effects is done under simplifying assumption that the only light degrees of freedom are the SM particles and the leptoquark $\Delta$.

The angle $\xi$ that enters $U_R=(O_2(\xi)\,O_3(\phi)\,O_1(\theta))^T$ cannot be deduced using the constraints imposed by the form of $Z$. We thus resort to proton decay signatures to see whether it could be fixed if and when proton decay is observed. We assume that proton decay is dominated by processes that involve exchange of gauge boson leptoquarks.

To predict gauge boson mediated proton decay signatures for two-body final states with good accuracy one needs to know all unitary transformations in the charged fermion sector, masses of all proton decay mediating gauge bosons and a gauge coupling constant. In $SU(5)$ the masses of LQ gauge bosons coincide with the scale of unification, i.e., GUT scale $m_{\mathrm{GUT}}$ and the gauge coupling at that scale is $\alpha_{\mathrm{GUT}}=g^2_\mathrm{GUT}/(4 \pi)$. Proton decay predictions in our setup hence depend on $m_{\mathrm{GUT}}$, $\alpha_{\mathrm{GUT}}$ and angle $\xi$. (What actually enters decay widths is ratio $\alpha_{\mathrm{GUT}}/m^2_{\mathrm{GUT}}$ and $\xi$.)

\begin{figure}[t]
\begin{center}
\includegraphics[width=11cm]{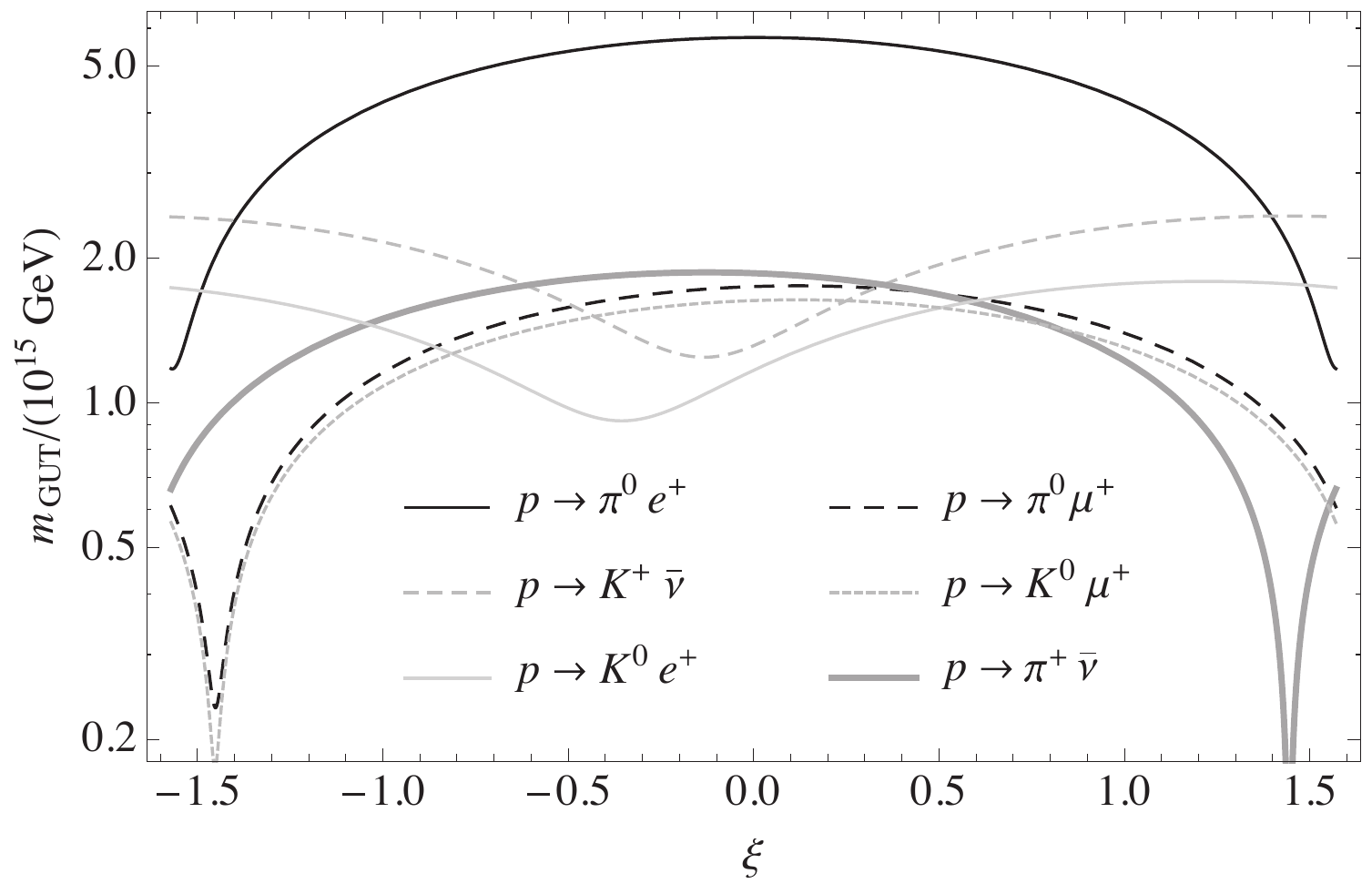}
\end{center}
\caption{Limits on the unification scale $m_\mathrm{GUT}$ as a function of angle $\xi$ as inferred from the latest experimental constraints on two-body proton decay. The region below each curve represents parameter space that is excluded by corresponding experimental limit.}
\label{fig:1}
\end{figure}

We can turn the argument around and use experimental limits on partial proton decay lifetimes to constrain the scale of unification. We present limits on $m_\mathrm{GUT}$ as a function of $\xi$ in Fig.~\ref{fig:1}. The region below each curve is excluded by current experimental limit on corresponding process. To generate results shown in Fig.~\ref{fig:1} we use the following experimental input on partial proton decay lifetimes: ${\tau}_{p \rightarrow \pi^0 e^+}> 1.3 \times 10^{34}$\,years~\cite{Nishino:2012ipa}, ${\tau}_{p \rightarrow K^+ \bar{\nu}}>  4.0 \times 10^{33}$\,years~\cite{Miura:2010zz}, ${\tau}_{p \rightarrow K^0 e^+}>  1.0 \times 10^{33}$\,years~\cite{Kobayashi:2005pe}, ${\tau}_{p \rightarrow \pi^0 \mu^+}>  1.1 \times 10^{34}$\,years~\cite{Nishino:2012ipa}, ${\tau}_{p \rightarrow K^0 \mu^+}>  1.6 \times 10^{33}$\,years~\cite{Regis:2012sn} and ${\tau}_{p \rightarrow \pi^+ \bar{\nu}}>  3.9 \times 10^{32}$\,years~\cite{Abe:2013lua}. We take $\hat \alpha=-0.0112$\,GeV$^3$~\cite{Aoki:2008ku}, where $\hat \alpha$ is the relevant nucleon matrix element. For the strength of unified gauge coupling we take $\alpha_{\mathrm{GUT}}=0.033$ and for the leading-log renormalization corrections of the relevant $d=6$ operator coefficients we use $A_{S\,L} = 2.6$ and $A_{S\,R} = 2.4$~\cite{Dorsner:2009mq}. These values vary very slightly with a change in the field content of the particular $SU(5)$ scenario in a framework without supersymmetry. For example, the model with $15$-dimensional scalar representation yields $\alpha_{\mathrm{GUT}}=0.031$, $A_{S\,L} = 2.8$ and $A_{S\,R} = 2.6$. The dependence of the decay widths on unitary transformations is taken from Ref.~\cite{FileviezPerez:2004hn} and the setup on how to propagate $A_{S\,L}$ and $A_{S\,R}$ coefficients from the GUT scale down to the low-energy scale is described in Ref.~\cite{Dorsner:2009mq}.  Note that the lifetimes scale with $m_{\mathrm{GUT}}$ ($\alpha_{\mathrm{GUT}}$) to the fourth (second) power.

The reasons we present these predictions for proton decay are twofold. Firstly, this clearly demonstrates the level of predictivity of our ansatz. Secondly, and more importantly, this demonstrates that even the simplest possible ansatz can predict that for some portions of parameter space it is not $p \rightarrow \pi^0 e^+$ channel that dominates if one assume that the main contribution to proton instability comes from the gauge boson exchanges. It is clear from Fig.~\ref{fig:1} that $p \rightarrow \pi^0 e^+$ is suppressed with respect to $p \rightarrow K^+ \bar{\nu}$ and even $p \rightarrow K^0 e^+$ in some parts of available parameter space, contrary to what is commonly stated in the literature. We can also see in Fig.~\ref{fig:1} that the proton decay signature for $p \rightarrow \pi^+ \bar{\nu}$ process is completely rotated away for one particular value of $\xi$. The possibility that this sort of suppression od individual decay modes could take place has been discussed before~\cite{Dorsner:2004jj,Dorsner:2004xa}. These findings only confirm that proton decay represents fertile, yet treacherous, ground to test Yukawa sector of the theory~\cite{DeRujula:1980qc}. 

Our analysis is performed for the proton decay signatures due to a tree-level exchange of gauge bosons. We are potentially in a position to present similar analysis for the two-body proton decay signatures due to scalar exchange. For example, we know how proton mediating leptoquarks in $45$-dimensional representation couple to the down-type quark and the charged lepton sectors, up to angle $\xi$ and the overall scale parameter $v_{45}$, i.e., the VEV of $\mathbf{45}$. We accordingly point out one particular property of our ansatz with regard to the proton decay signatures through scalar LQ exchange. Namely, there exists a color triplet leptoquark $(3,1)_{-1/3}$ in $\mathbf{45}$ that is synonymous with matter instability as can be seen from Tab.~\ref{tab:states}. But, in this particular setup, all $d=6$ proton decay operators due to exchange of that triplet are completely suppressed. (Relevant coefficients of $d=6$ operators for the color triplet are given in Eqs.~(10), (11) and (12) of Ref.~\cite{Dorsner:2012nq}.)   

To complete our numerical study we spell out products $v_{45} z_{2i}$, $i=1,2,3$, at the GUT scale: $v_{45} z_{21}=0.14\e{GeV}$, $v_{45} z_{22}=0.17\e{GeV}$ and $v_{45} z_{23}=0.48\e{GeV}$. $V_\mrm{PMNS}$ entries that are needed to convert $\tilde z_{2i}$ into $z_{2i}$ are adopted from Ref.~\cite{GonzalezGarcia:2012sz}. Of course, the values we present should be run down from the GUT scale to the electroweak scale where we generate constraints on $\tilde z_{2i}$, $i=1,2,3$, coefficients. We estimate that the low-energy values will be increased by a scaling factor $f_\mathrm{RGE}$ that is within the following range: $f_\mathrm{RGE} \in [1.1,3.7]$~\cite{Dorsner:2011ai}. This holds as long as the initial value at the low energy scale for $|\tilde z_{2i}|$, $i=1,2,3$, is below $0.5$. The ratio between $\tilde z_{2i}$, $i=1,2,3$, coefficients, however, should be constant with regard to the running effects.

\section{GUT connection between $\ell \to \ell' \gamma$ and $B \to
  D^{(*)} \tau \bar \nu$}
\label{FEEDBACK}
We may take advantage of the known hierarchy in how $\Delta^{(5/3)}$ and charm quark couple to charged leptons. Recall that the
CKM induced couplings to $u_L$, $c_L$ and $t_L$ are
hierarchically suppressed, due to our ansatz of Eq.~\eqref{eq:Yuk23}, with factors $V_{ub}$, $V_{cb}$ and $V_{tb}$,
respectively. Similar hierarchy in the $Z$ matrix
originates, on the other hand, from the requirement of realistic
fermion masses in the GUT setting. In turn, this reduces the parameter
set of the model to two independent Yukawa couplings. One of these
must be $y_{33}$ while the other can be any one of $z_{2j}$ or $\tilde
z_{2j}$, $j=1,2,3$. We choose $\tilde z_{22}$, the coupling of
$\Delta^{(5/3)}$ scalar to $c \mu$ pair, to be the other independent parameter while remaining $z_{2j}$ and $\tilde
z_{2j}$ are determined
either from the hierarchical pattern~\eqref{eq:zhier} or via PMNS
relation~\eqref{eq:Yuk23}.

The PMNS rotation connecting $\tilde z_{2j}$ and $z_{2k}$ couplings
reduces to a simple linear relation between $\tilde z_{22}$ and
$z_{23}$ when Eq.~\eqref{eq:zhier} is applied:
\begin{equation}
\label{eq:a}
 z_{23} =   \tilde z_{2k} V_{k3} \approx \tilde z_{22} \, c_{13}( s_{23} + 3.22\, c_{23})\,.
\end{equation}
The numerical factor $3.22$ in Eq.~\eqref{eq:a} comes from the hierarchy between $\tilde
z_{2k}$ couplings. $V_{ij}$, $s_{ij}$ and $c_{ij}$ denote the PMNS
matrix elements and the mixing angles that parameterize it,
respectively. Using the $3\,\sigma$ ranges for mixing angles from a recent
PMNS fit~\cite{GonzalezGarcia:2012sz} we find
\begin{equation}
 z_{23} =  \omega \tilde z_{22}\,,\qquad 2.63 < \omega < 3.17\,.
\end{equation}

\begin{figure}[!htcb]
  \centering
  \includegraphics[width=.7\textwidth]{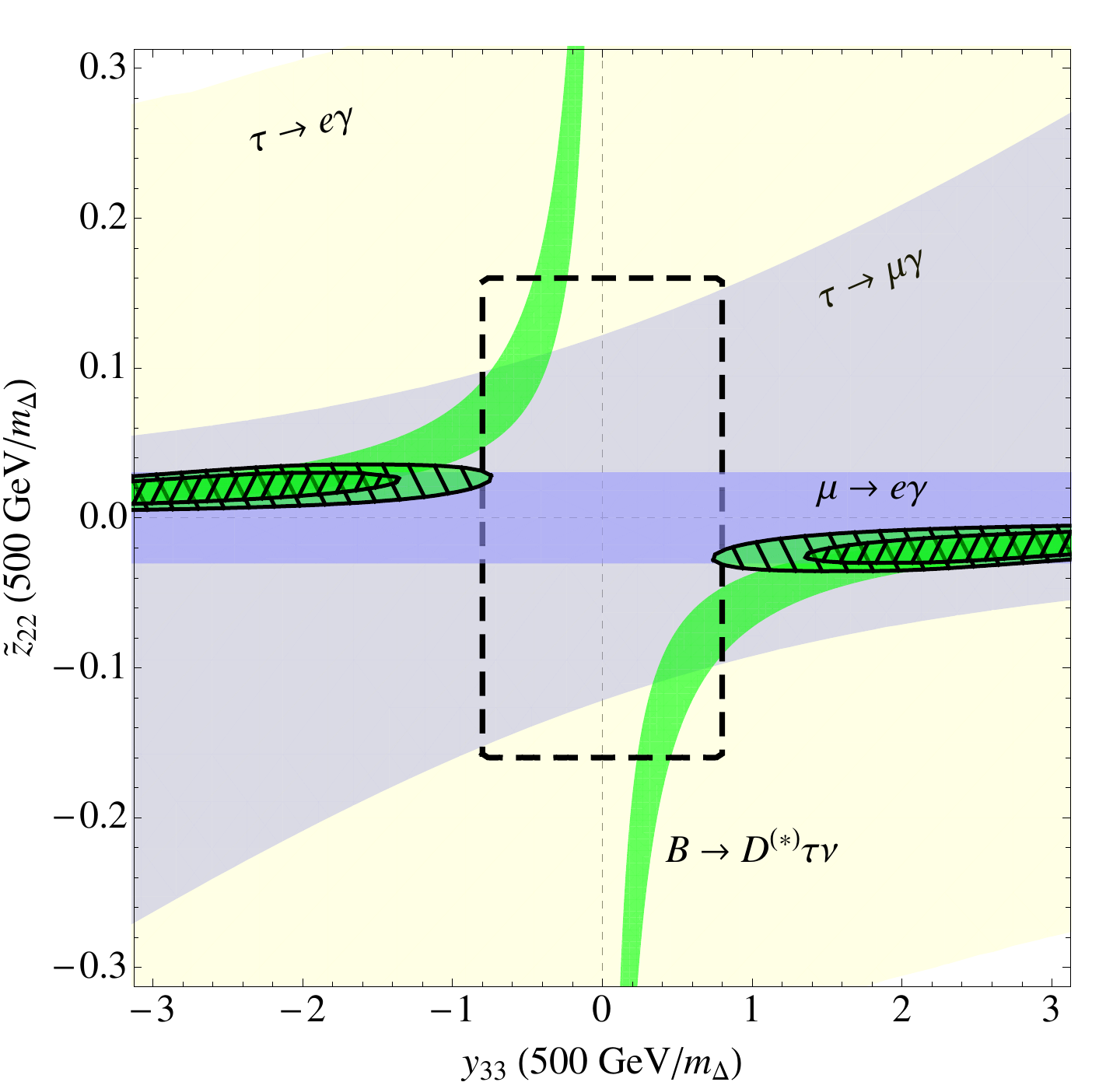}
  \caption{Constraints on the couplings to $b\tau$ ($y_{33}$) and to
    $c\mu$ ($\tilde z_{22}$) coming from the $1\,\sigma$ region of
    $\mc{R}^{(*)}_{\tau/\ell}$ (thin hyperbolic region), $90\,\%$ CL
    upper bounds on $\mu \to e \gamma$, $\tau \to \mu \gamma$ and $\tau \to
    e\gamma$. Dashed frame represents
    the region where couplings remain perturbative all the way to the GUT scale, as
    explained in the text. Doubly (singly) hatched area is
    allowed at $1\sigma$ ($2\sigma$).}
  \label{fig:constraints}
\end{figure}
Effect of the aforementioned experimental constraints on $(3,2)_{7/6}$
with the minimal Yukawa texture~\eqref{eq:Yuk23}, additionally
restricted by the pattern of fermion masses, is shown on
Fig.~\ref{fig:constraints}. As expected, the central role is played by
the constraint on $g_S$, although $\tau \to \mu \gamma$ reduces the
parameter space remarkably. Due to suppressed coupling to $e$,
sensitivity of $\tau \to e \gamma$ and $\mu \to e \gamma$ observables
is reduced, however, the latter overcomes this suppression by a very
stringent experimental upper bound and therefore has the most important role
next to constraint on $g_S$. An order of magnitude improvement on the experimental
bound on $\mu \to e \gamma$ would cause tension with the
$R(D^{(*)})$ observables, and smaller values of
the $g_S$ coupling would be preferred. 
Note that only the $2\,\sigma$ region (hatched in Fig.~\ref{fig:constraints}) is
overlapping with the region where $y_{33}$ is perturbative all the way
to the GUT scale. We mention in passing that for mass $200\e{GeV} <
m_\Delta < 1\e{TeV}$ all predictions are approximately invariant under
rescaling of couplings and $m_\Delta$ by same factor, as indicated on
the axes of Fig.~\ref{fig:constraints}.

The $2\sigma$ region at $m_\Delta = 500\e{GeV}$ in the
  $y_{33}-\tilde z_{22}$ plane implies the following bounds
\begin{equation}
\label{eq:b}
 |y_{33}| > 0.74\,,\qquad |\tilde z_{22}| < 0.037\,.
\end{equation}
Region that satisfies perturbativity of the couplings all the way to the GUT
scale, outlined by dashed frame in Fig. 6, restricts the above 
$2\sigma$ ranges to
\begin{equation}
\label{eq:c}
 0.74 < |y_{33}| < 0.80\,,\qquad 0.021 <|\tilde z_{22}| < 0.032\,.
\end{equation}
These bounds can be further put to use to generate limits on allowed values of $v_{45}$. We find, using the GUT deduced values for the product $v_{45} \tilde
z_{2j}$, $j=1,2,3$, that 
\begin{equation}
\label{eq:d}
f_{\mathrm{RGE}}\,5.0\,\mathrm{GeV}  < v_{45} < f_{\mathrm{RGE}}\,7.6\,\mathrm{GeV}\,.
\end{equation}
This, in turn, confirms that the GUT embedding is self-consistent as the required values for $v_{45}$ lead to small radiative corrections to Yukawa couplings.

\section{Predictions}
\label{PREDICTIONS}
\subsection{$B_c \to \tau\nu$}
The leptonic decay is governed by the same effective Hamiltonian as
semileptonic decay and is therefore directly related to the latter,
remaining insensitive to the underlying correlations between the LQ
couplings. The only difference with respect to the semileptonic decay
is that the decay $B_c\to\tau\nu_\tau$ probes only the (pseudo)scalar
operator of effective Hamiltonian~\eqref{Hamiltonian} and is
insensitive to tensor one due to vanishing matrix element $\langle
0|\bar{c}\sigma_{\mu\nu}b|B\rangle$.  The value of decay constant of
$B_c$ meson, $f_{B_c}=0.427(6)(2)\e{GeV}$ has been recently calculated
by HPQCD Collaboration in lattice QCD with fully relativistic
formalism~\cite{HPQCD}.  The branching ratio of the process in the SM
is calculated using the formula
\begin{equation}
\begin{split}
\mathcal{B}(B_c\to\ell\nu)=\frac{m_{B_c}}{8\pi}\tau_{B_c} f_{B_c}^2|G_F V_{cb}m_\ell|^2\Big(1-\frac{m_\ell^2}{m_{B_c}^2}\Big)^2 r^2\,,
\end{split}
\end{equation}
where  factor $r$ given by
\begin{equation}
r=\Bigg |1+\frac{m_{B_c}^2}{m_\tau(m_b+m_c)}g_S\Bigg |\,.
\end{equation}
Here $\tau_{B_c}$ is lifetime of $B_c$ meson and $g_S$ is the Wilson
coefficient defined in Eq.~\eqref{gS}. The HPQCD Collaboration gives
SM prediction $\mc{B}(B_c\to\tau\nu)=0.0194(18)$~\cite{HPQCD}. For the
best fit value $g_S=-0.37$ the branching ratio is suppressed by the
factor $r^2\simeq 0.36$ with respect to SM, while also large
enhancement ($r^2\simeq 84$) is allowed for $g_S\simeq
1.8\pm0.4\,i$. The possible large enhancement by the same operator, in
the context of Two Higgs Doublet models was also recently reported
in~\cite{Pich-Celis}. One should note that the production rate of the
$B_c$ meson at high energy colliders is several orders of magnitude
smaller than for other $B$ mesons~\cite{LHCb-Bc:2012}, and the
observation prospects for $B_c$ leptonic channels are not very
promising.

\subsection{$t \to c \tau^+ \tau^-$}
\begin{figure}[!hctp]
  \centering
      \includegraphics[width=0.5\textwidth]{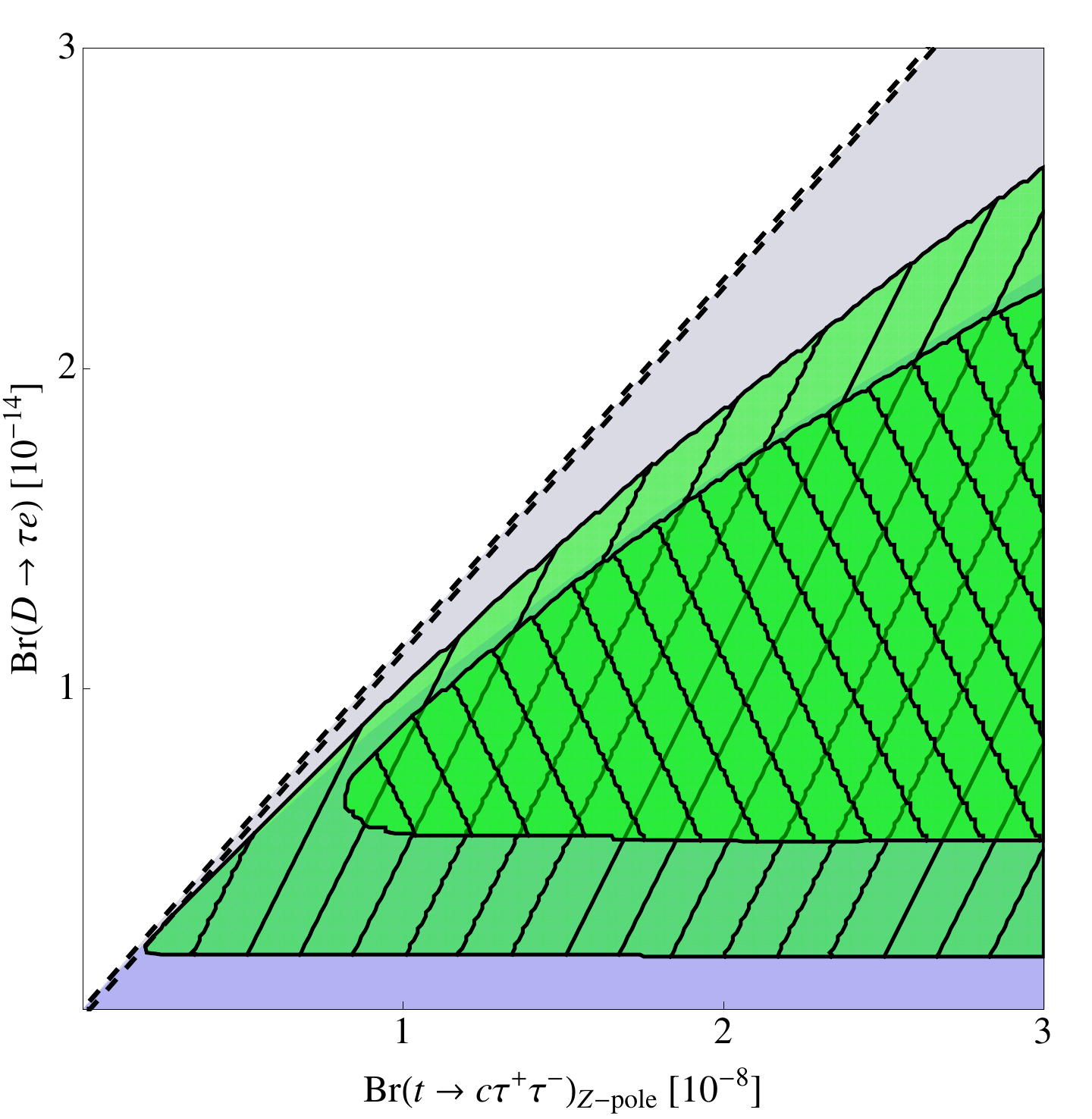}
  \caption{Predictions of the branching fractions of $t \to c \tau^+
    \tau^-$ under the $Z \to \tau^+ \tau^-$ peak
    ($|m_{\tau\tau}^2-m_Z^2| < m_Z \Gamma_Z$) and $\bar D^0 \to \tau^-
    e^+$. The dashed line shows where the perturbativity to the GUT
    scale holds, whereas the color coding follows Fig.~\ref{fig:constraints}.
}
   \label{fig:topD}
\end{figure}

Experiments at LHC have already established upper bounds for events
with rare decays $t \to q Z$ with $q=u,c$ and with $Z$ reconstructed
from light leptons. In the model with colored scalar a $t$-channel
exchange of $\Delta^{(5/3)}$ contributes to the $t \to c \tau^+
\tau^-$ decay.  Tau leptons may further decay to light leptons and
feed into the SM signal of $t \to Z c \to \ell^+ \ell^- c$, or they
could be reconstructed in a study dedicated to the $\tau^+ \tau^-$
mode.  The decay $t \to c \tau^+ \tau^-$ can be treated in an
effective approach, using a Lagrangian that encompasses the couplings
induced by the $\Delta^{(5/3)}$ exchange~\eqref{eq:L53},
\begin{equation}
  \label{eq:Lefftop}
  \mc{L}_\mrm{eff} = A (\bar\tau_R \gamma^\mu \tau_R) (\bar c_L
  \gamma_\mu t_L) + B \left[ (\bar\tau_R \tau_L) (\bar c_R  t_L) +
    \frac{1}{4}  (\bar\tau_R \sigma^{\mu\nu}\tau_L) (\bar c_R
    \sigma_{\mu\nu}  t_L)  \right]\,.
\end{equation}
Production mechanism of $t$ quarks may also be affected by colored
scalars and can be treated independently of the decay branching
fractions. The differential decay width of the $t \to c \tau^+ \tau^-$
decay expressed in terms of normalized tau-pair invariant mass, $\hat s \equiv m_{\tau\tau}^2/m_t^2$, reads
\begin{equation}
\frac{d\Gamma}{d \hat s}= \frac{m_t^5\left(1-\hat s\right)}{3072 \pi^3}
\left[48  |A|^2\,\hat s\left(1-\hat s\right)+|B|^2 \left(11+20 \hat s 
    -13 \hat s^2\right)\right]\,.
\end{equation}
A very small SM contribution to this decay has been neglected
altogether~\cite{Glover:2004cy}.  The two Wilson coefficients,
given in terms of the couplings in the renormalizable
Lagrangian~\eqref{eq:L53}, are
\begin{equation}
  \label{eq:tAB}
  A = -\frac{|y_{33}|^2 V_{cb} V_{tb}^*}{2m_\Delta^2}\,,\qquad B =
  \frac{y_{33} \tilde z_{23} V_{tb}^*}{2m_\Delta^2}\,.
\end{equation}
The values of the above couplings are constrained by $B \to D^{(*)}
\tau \bar \nu$, $\tau \to \mu \gamma$, and perturbativity requirement
that are mapped to the plane of predictions of $t \to c \tau^+ \tau^-$
and $\bar D^0 \to \tau^- e^+$ in Fig.~\ref{fig:topD}, where the
branching fraction of top decay below the $Z$ peak,
$|m_{\tau\tau}^2-m_Z^2| < m_Z \Gamma_Z$, is expected to lie in the
range of few times $10^{-9}$. Relaxing the perturbativity constraint
can give an order of magnitude enhancement.  We have also checked that
the $t \to c \tau^+ \tau^-$ total branching fraction, that is not
limited to the region where $m_{\tau\tau}^2 \approx m_Z^2$, is
further enhanced by one order of magnitude.

\subsection{$\bar D^0 \to \tau^- e^+$}
This is the only allowed lepton flavor violating decay of the neutral charm
meson since the decay to $\tau^{\mp} \mu^\pm$ is kinematically closed. The
effective Lagrangian is a combination of scalar and tensor parts due to
Fierz identities
\begin{equation}
\mc{L}_\mrm{eff} = -\frac{V_{ub}^* y_{33} \tilde z_{21}}{2m_\Delta^2} \,\left[(\bar
  \tau_R e_L)(\bar c_R   u_L) + \frac{1}{4} (\bar
  \tau_R \sigma^{\mu\nu} e_L)(\bar c_R \sigma_{\mu\nu}  u_L) \right]\,,
\end{equation}
and only the scalar operator contributes to the decay width. The decay
width is suppressed by small coupling to $e$, tiny phase space
($m_\tau/m_{D^0} = 0.95$), and $V_{ub}$:
\begin{equation}
  \label{eq:BrDtaue}
  \mc{B}(\bar D^0 \to \tau^- e^+) = \frac{\tau_{D^0}}{256\pi}
  \frac{m_D^2}{m_c^2} \left(1-m_\tau^2/m_D^2\right)^2\,m_D^3 f_D^2\,|V_{ub}|\, \left(\frac{|\tilde z_{21}y_{33}|}{m_\Delta^2}\right)^2\,.
\end{equation}
The combined effect of all suppression factors renders the branching
fraction of this decay in the ballpark of $10^{-15}$ if
perturbativity to the GUT scale is required (see
Fig.~\ref{fig:topD}). Relaxing this criterion can increase branching
fraction up to 1 order of magnitude.

\section{Conclusions}
\label{CONCLUSIONS}
In order to explain the observed deviation from the Standard Model
prediction of the ratios $R(D)$ and $R(D^{*})$ we have explored the
possibility of introducing a single light leptoquark state. Of all possible scalar and vector
states with renormalizable couplings we have identified a scalar leptoquark
with the Standard Model quantum numbers $(3,2)_{7/6}$ as the most
suitable one. 

In the framework of effective theory for semileptonic decays the
tree-level exchange of charge-$2/3$ component of this leptoquark introduces a new operator, a
particular combination of scalar and tensor currents, that interferes
constructively with the vector current of the Standard Model. We have
endowed the leptoquark with a minimal set of flavor couplings that are
adequate to explain the discrepancy in $b \to c \tau \bar\nu$
processes and do not disturb flavor changing processes
involving first two generation of quarks and leptons. A combination of
$b \tau$ and $c \nu$ Yukawa couplings suffices to explain the
anomalies in $B \to D \tau \bar \nu$ as well as in $B \to D^* \tau
\bar \nu$ decays. As it turns out, the latter decay has dominant
sensitivity to the tensor operator, although the scalar operator is
almost an order of magnitude larger. The overlap of the two
observables prefers small effective coupling that interferes
positively with the Standard Model.

Regardless of the careful choice of the Yukawa couplings for the
charge-2/3 component of this leptoquark, the charge-5/3 component will
nevertheless induce lepton and quark flavor changing processes. We
analyze in detail the constraints imposed by $Z \to b\bar b$ decay,
value of the muon magnetic moment, lepton flavor violating decays $\mu
\to e \gamma$, $\tau \to \mu \gamma, e \gamma$, and $\tau$ electric
dipole moment. All these occur at one-loop level. The long standing
anomaly in $Z \to b \bar b$ requires deviation of both couplings $g_R$
and $g_L$. Since the leptoquark $(3,2)_{7/6}$ connects in vertices to
$b_L$, one can modify only the left handed coupling. This bound gives
us a constraint on the coupling $y_{33}$, i.e., to $b\tau$ pair, that
is weaker than perturbativity limit on $y_{33}$. Presence of this
leptoquark pushes the Standard Model prediction of muonic $g-2$
further away from the measured value. The experimental result puts a
quite strong constraint on the coupling to $c\mu$, $\tilde z_{22} \le
0.51$, for the typical mass of $m_\Delta =500$\,GeV in a simple
  extension of the SM with the aforementioned LQ.  In the case of
electric dipole moment, CP violating product of two different
couplings can arise only in the case of $\tau$ lepton. Current
experimental bound from Belle experiment is orders of magnitude too
weak to address the problem of phase of the product $V_{cb} y_{33}^*
\tilde z_{23}^*$, at least in the perturbative setting. On the other
hand, current experimental bound on the lepton flavor violating decays
$\tau \to \ell \gamma$ with $\ell = \mu, \tau$, and especially $\mu
\to e \gamma$, provide a stringent constraint on the underlying
leptoquark couplings to $\tau \bar c$, $\mu \bar c$, and $e \bar c$.

After completing the analysis of phenomenological constraints on
relevant leptoquark couplings we consider interplay between these constraints
and the mass generation mechanism for charged leptons and down-type
quarks in a GUT framework. We find that the minimal set of Yukawa
couplings is not only compatible with the $SU(5)$ unification, a
natural environment for colored scalars, but specifies all matter
mixing parameters except for one angle in the up-type quark sector. We
present predictions for the proton decay signatures through gauge
boson exchange, as a function of that angle, and show that $p
\rightarrow \pi^0 e^+$ process is suppressed with respect to $p
\rightarrow K^+ \bar{\nu}$ and even $p \rightarrow K^0 e^+$ in some
parts of available parameter space. This goes against the expectations
commonly encountered in the literature. The ansatz yields ratios
between the leptoquark couplings that mimics mass hierarchy in the down-type
quark and the charged lepton sectors. Namely, we find that $ \tilde
z_{21} : \tilde z_{22} : \tilde z_{23} = 0.024 : 0.32 : 1$.
In order to preserve perturbativity at the GUT scale we require that $|y_{33}|<0.8$
 and $|\tilde z_{23}|<0.5$ at the low energy scale. In this regime the radiative corrections to our 
ansatz for Yukawa structure are under control.
  
Both the approaches from the low energy phenomenology point of view
and the constraints coming from the GUT setup and
proton decay result in a number of predictions. An obvious independent
test of the $b \to c \tau \bar\nu$ process would be the decay $B_c \to
\tau \bar\nu_\tau$ that probes the same couplings as the semileptonic
decay. Presence of scalar operator in the effective Lagrangian entails
enhancements of 1 to 2 orders of magnitude of $B_c \to \tau
\bar\nu_\tau$ decay width, although an experimental search for this decay
is notoriously difficult. When the scalar state $(3,2)_{7/6}$ is a
part of $45$-dimensional GUT representation there are only two
independent couplings to consider. Then it turns out that $\tau \to
\mu \gamma$ decay should lie within one or two orders of magnitude
below the current experimental bound, and that $\mu \to e \gamma$
should be virtually one step beyong the reach of the MEG experiment bound, if this leptoquark is to explan
the measured values of $\cal R_{\tau/\ell}^{(*)}$. We make definite
predictions, within the $SU(5)$ setup, of the lepton flavor violating
process $\bar D^0 \to \tau^- e^+$ and of the rare process $t \to c
\tau^+ \tau^-$. The former decay is very suppressed. The top decay
with $\tau^+ \tau^-$ emulating the $Z \to \tau^+ \tau^-$ is also
expected to proceed with branching fraction less than
$10^{-8}$. However, a dedicated search for $t \to c \tau^+ \tau^-$ with
integrated ditau spectrum could find a branching fraction of up to $10^{-7}$.
We have also found allowed values for the vacuum
expectation value of the $45$-dimensional representation to be
$f_{\mathrm{RGE}}\,5.0\,\mathrm{GeV} < v_{45} <
f_{\mathrm{RGE}}\,7.6\,\mathrm{GeV}$, where $f_\mathrm{RGE} \in [1.1,3.7]$.

Due to the peculiar couplings of the $(3,2)_{7/6}$ leptoquark to
$\tau b$ pair and allowing it to couple to charm quark and all three
neutrinos we can successfully explain the anomalies observed
in semileptonic $b \to c$ transitions. If the presence of this state
indeed solves these anomalies then it should also be observed eventually
in lepton flavor violating decays of the tau leptons and muons and rare decays of
top quark to charm quark and a pair of tau leptons.

\acknowledgments

We acknowledge very useful discussions with D.~Be\v
  cirevi\' c and J.~F.~Kamenik. I.~D.\ acknowledges the SNSF support
  through the SCOPES project No.\ IZ74Z0\_137346. This work is
  supported in part by the Slovenian Research Agency.

\appendix

\section{Parameterization of the $B \to D^{(*)}$ form factors}
\label{sec:appBD}
For the $B \to D$ vector form factor we use expansion of
the function $G_1(w)$~\cite{CLN}
\begin{equation}
G_1(w)=G_1(1)[1-8\rho_ D^2 z(w)+(51\rho_D^2-10)z(w)^2-(252\rho_D^2-84)z(w)^3],
\end{equation}
in terms of variable
$z(w)=\frac{\sqrt{w+1}-\sqrt{2}}{\sqrt{w+1}-\sqrt{2}}$. The slope
parameter has been determined by Heavy Flavour Averaging Group~\cite{HFAG},
and has value $\rho_D^2=1.186 \pm 0.055$.

For the amplitudes of the $B \to D^{*} \ell \bar\nu$ decay, Eq.~\eqref{eq:BDstAmp}, the
hadronic helicity amplitudes are defined as contractions of matrix
elements of corresponding leptonic and hadronic currents with the
helicity four-vectors $\tilde{\epsilon}_\mu(\lambda)$:
\begin{subequations}
\begin{align}
& H_{V-A,\lambda}^{\lambda_{D^*}}(q^2,\cos\theta _\ell)= \tilde{\epsilon}_\mu(\lambda)\langle D^*(p_{D^*}),\epsilon(\lambda_{D^*})|\bar{c}\gamma^\mu(1-\gamma_5)b|B(p_B)\rangle\,,\\
& H_{P}^{\lambda_{D^*}}(q^2,\cos\theta_\ell)= \langle D^*(p_{D^*}),\epsilon(\lambda_{D^*})|\bar{c}(1-\gamma_5)b|B(p_B)\rangle\,,\\
& H_{T,\lambda\lambda'}^{\lambda_{D^*}}(q^2,\cos\theta_\ell) =\tilde{\epsilon}_\mu(\lambda)\tilde{\epsilon}_\nu(\lambda')\langle D^*(p_{D^*}),\epsilon(\lambda_{D^*})|\bar{c}\sigma^{\mu\nu}(1-\gamma_5)b|B(p_B)\rangle. \label{helicity-amplitudes}
\end{align}
\end{subequations}
The explicit form of these functions is found in the literature, see e.g.~\cite{korner-schuler1,korner-schuler2,M.Tanaka2012,FKN}.
Leptonic helicity amplitudes are defined in analogous fashion.

Functions $R_i(w)$ entering form factors $A_{1,2,3}(w)$ and $V(w)$ are
expanded around point $w=1$~\cite{CLN,FKN}
\begin{equation}
\begin{split}
&h_{A_1}(w)=h_{A_1}(1)[1-8\rho^2 z(w) +(53\rho^2-15)z(w)^2-(231-91)z(w)^3]\,,\\
&R_0(w)=R_0(1)-0.11(w-1)+0.01(w-1)^2\,,\\
&R_1(w)=R_1(1)-0.12(w-1)+0.05(w-1)^2\,,\\
&R_2(w)=R_2(1)+0.11(w-1)-0.06(w-1)^2\,,\\
&R_3(w)=R_3(1)-0.052(w-1)+0.026(w-1)^2.\label{Rs}\\
\end{split}
\end{equation}
The numerical values of $R_i(w)$ used in calculations are $h_{A_1}(1)=
0.919 \pm 0.035$, 
$R_0(1)=1.14$, $R_1(1)=1.403\pm 0.033$, $R_2(1)=0.854\pm 0.020$ and
$R_3(1)=1.22$~\cite{Belle2010,FKN}.

\section{EW gauge couplings of the scalar $(3,2)_{7/6}$}
\label{sec:Zbbapp}
Consistently with Eq.~\eqref{eq:Zbbdef} we define the 
covariant derivative as $D_\mu = \partial_\mu -i g W_\mu^3
\frac{\sigma^3}{2} - i g^\prime B_\mu Y$ and the
weak mixing angle as $W_\mu^3 = \cos \theta_W Z_\mu + \sin \theta_W
A_\mu$. The Feynman rules for colored scalar-gauge boson interactions
are listed in Tab.~\ref{tab:rules}.
\begin{table}[!hctp]
  \centering
  \begin{tabular}{|c|c|}
\hline
    \includegraphics[width=0.27\textwidth]{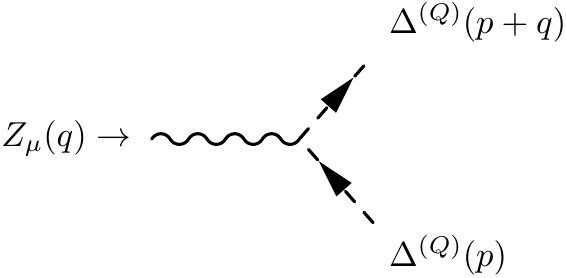} & \includegraphics[width=0.27\textwidth]{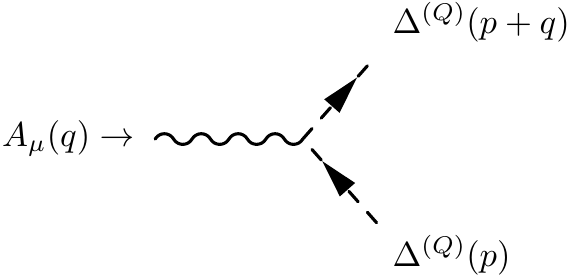} \\&\\
 $-ig \left(-Q c_W + \frac{Y}{c_W}\right) (2 p + q)_\mu
    \epsilon_Z^\mu$&
     $ieQ (2 p + q)_\mu
    \epsilon_A^\mu$\\
\hline
  \end{tabular}
 \caption{Feynman rules for scalar couplings to neutral gauge bosons.}
 \label{tab:rules}
\end{table}

\section{Loop function of $Z \to b \bar b$}
\begin{align}
g_0(x) &=-12 x^2 \Bigg[-\mrm{Li}_2\Big[(1/x-1/x^2) (x-i \sqrt{x-1/4}-1/2)\Big]-\mrm{Li}_2\Big[(1/x-1/x^2) (x+i
   \sqrt{x-1/4}-1/2)\Big]\\
&\phantom{=}+\mrm{Li}_2\Big(1-1/x\Big)
 +\mrm{Li}_2\Big[(x-i
   \sqrt{x-1/4}-1/2)/x\Big]+
\mrm{Li}_2\Big[(x+i
   \sqrt{x-1/4}-1/2)/x\Big]
\Bigg]\nn\\
&\phantom{=}+2 \Big[\pi ^2 x^2+3 i \pi  \sqrt{1-4 x} (2 x-1)+\sqrt{1-4
   x} (2 x-1) \log (8)\Big]\nn\\
&\phantom{=}+3 (4 x-5)+6 \sqrt{1-4 x} (1-2 x) \log
\left(\frac{-2 x+\sqrt{1-4 x}+1}{x}\right)\,,\nn\\
g_2(x)&=16x^2\Bigg[\frac{3}{2} \mrm{Li}_2\Big[1+\frac{1}{x}\Big]\\
&\phantom{=}+\mrm{Li}_2\left((1/x-1/x^2) \left(x-i
   \sqrt{x-\frac{1}{4}}-\frac{1}{2}\right)\right)+\mrm{Li}_2\left((1/x-1/x^2) \left(x+i
   \sqrt{x-\frac{1}{4}}-\frac{1}{2}\right)\right)\nn\\
&\phantom{=}-\mrm{Li}_2\left(\frac{x-1}{x}\right)-\mrm{Li}_2\left(\frac{x-i
   \sqrt{x-\frac{1}{4}}-\frac{1}{2}}{x}\right)-\mrm{Li}_2\left(\frac{x+i
   \sqrt{x-\frac{1}{4}}-\frac{1}{2}}{x}\right)\Bigg]\nn\\
&\phantom{=}-4 \pi ^2 x^2/3+48 i \pi  x^2 \log
   \left(\frac{1}{x}+1\right)+4 i \pi  \left(2 \sqrt{1-4 x}-3\right)
   (2 x-1)-2 (4 x+7)\nn\\
&\phantom{=}+8 \sqrt{1-4 x} (1-2 x) \log \left(\frac{-2 x+\sqrt{1-4 x}+1}{x}\right)+12 (1-2 x) \log (x)+8
   \sqrt{1-4 x} (2 x-1) \log (2)\,.  \nn
\end{align}

\bibliography{citations}

\end{document}